\documentclass[aps,twocolumn,floatfix,letterpaper,nofootinbib]{revtex4-2}

\usepackage[dvipsnames]{xcolor}

\newcommand{\gSq}{\mathbb{Sq}}

\newcommand{\RZ}{{\mathbb{R}/\mathbb{Z}}}
\newcommand{\QZ}{{\mathbb{Q}/\mathbb{Z}}}
\newcommand\se[1]{\overset{\scriptscriptstyle #1}{=}}

\newcommand\hcup[1]{\underset{{\scriptscriptstyle #1}}{\smile}}
\newcommand\toZ[1]{\lfloor #1 \rceil}

\newcommand{\hstar}{\mathop{*}} 
\def\wdg{{\mathchoice{\,{\scriptstyle\wedge}\,}{{\scriptstyle\wedge}}
{{\scriptscriptstyle\wedge}}{{\scriptscriptstyle\wedge}}}}

\newcommand{\p}{\partial}

\begin{document}

\begin{titlepage}

\title{Emergent Higher-Symmetry Protected Topological Orders\\
in the Confined Phase of $U(1)$ Gauge Theory}

\author{Salvatore D. Pace}
\affiliation{Department of Physics, Massachusetts Institute of Technology, Cambridge, Massachusetts 02139, USA}

\author{Xiao-Gang Wen} 
\affiliation{Department of Physics, Massachusetts Institute of Technology,
Cambridge, Massachusetts 02139, USA}

\date{\today}

\begin{abstract} 

We consider compact $U^\kappa(1)$ gauge theory in 3+1D with a general
$2\pi$-quantized topological term ${\sum_{I, J =1}^{\kappa} \frac{K_{IJ}}{4\pi}
\int_{M^4}F^I\wedge F^J}$, where $K$ is an integer symmetric matrix with even
diagonal elements and ${F^I = \mathrm{d} A^I}$.  At energies below the gauge
charges' gaps but above the monopoles' gaps, this field theory has an emergent
${\mathbb{Z}_{k_1}^{(1)} \times \mathbb{Z}_{k_2}^{(1)} \times \cdots}$
1-symmetry, where $k_i$ are the diagonal elements of the Smith normal form of
$K$ and $\mathbb{Z}_{0}^{(1)}$ is regarded as a $U(1)$ 1-symmetry.  In the
$U^\kappa(1)$ confined phase, the boundary can have a phase whose IR properties
are described by Chern-Simons field theory. Such a phase has a
${\mathbb{Z}_{k_1}^{(1)} \times \mathbb{Z}_{k_2}^{(1)} \times \cdots}$
1-symmetry that can be \textit{anomalous}.  To show these results, we develop a
bosonic lattice model whose IR properties are described by this continuum field
theory, thus acting as its UV completion.  The lattice model in the
aforementioned limit has an exact  ${\mathbb{Z}_{k_1}^{(1)} \times
\mathbb{Z}_{k_2}^{(1)} \times \cdots}$ 1-symmetry.  We find that the
short-range entangled gapped phase of the lattice model, corresponding to the
confined phase of the $U^\kappa(1)$ gauge theory, is a symmetry protected
topological (SPT) phase for the ${\mathbb{Z}_{k_1}^{(1)} \times
\mathbb{Z}_{k_2}^{(1)} \times \cdots }$ 1-symmetry, whose SPT invariant is
${\ee^{\ii \pi \sum_{I, J} K_{IJ} \int_{\cM^4} B_I  \smile B_J + B_I
\underset{{\scriptscriptstyle 1}}{\smile} \dd B_J} \ee^{\ii \pi  \sum_{I<
J}K_{IJ} \int_{\cM^4} \dd B_I \underset{{\scriptscriptstyle 2}}{\smile} \dd B_J
}}$.  Here, the background $\RZ$-valued 2-cochains $B_I$ satisfy ${\dd  B_I =
\sum_I B_{I} K_{IJ} = 0~\mod~1}$ and describe the symmetry twist of the
${\mathbb{Z}_{k_1}^{(1)} \times \mathbb{Z}_{k_2}^{(1)} \times \cdots }$
1-symmetry.  We apply this general result to a few examples with simple $K$
matrices.  We find the non-trivial SPT order in the confined phases of these
models and discuss its classifications using the fourth cohomology group of the
corresponding 2-group.

\end{abstract}

\maketitle

\end{titlepage}

\setcounter{tocdepth}{1} 
{\small \tableofcontents }

\section{Introduction}

Symmetry protected topological (SPT) phases of quantum matter are short-range
entangled gapped phases whose ground states cannot be adiabatically connected
to a trivial product state due to the presence of a
symmetry~\cite{CLW1141,CGL1204, S14054015, W161003911}.  The boundary of
a SPT phase is nontrivial because the symmetry is realized anomalously on the
boundary.  Since the SPT bulk has a trivial intrinsic topological order, the
boundary theory by itself (\ie without bulk) is perfectly consistent as a
lattice theory, and the 't Hooft anomaly ensures it cannot be in a trivial
phase. However, upon turning on background gauge fields of the symmetry, the
boundary theory is no longer a physical theory and can only exist as the
boundary of an invertible phase (i.e., an SPT). From the prospective of
anomaly inflow, the SPT order in the bulk provides a unique characterization of
the 't Hooft anomaly on the boundary.

Since their discovery, there have been numerous interesting generalizations of
SPT phases. This includes SPT phases with intrinsic topological order in the
bulk~\cite{LS0903, SB10051076, NS11063989, MQ11116816}, SPT phases with a
gapless bulk~\cite{PB09072962, SP170501557, TV200806638}, and higher-order SPTs
where edge states exist only on a subregion of the boundary~\cite{HH170509243,
YD180709788}. In this paper, we consider the generalization where the SPT order
is protected by a higher-symmetry~\cite{TK151102929,Y150803468,
BM170200868,KR180505367,WW181211967,W181202517,TW190802613, JW200900023}.

Global symmetries, called 0-symmetries, are symmetries whose transformation
acts on a codimension-1 submanifold of spacetime (e.g., all of space), and the
charged operators act on a single point in spacetime, creating a
$0$-dimensional object in space. Higher-symmetry is a generalization that
includes $p$-symmetry, where now the symmetry transformation acts on a
codimension-${(p+1)}$ submanifold of spacetime and the charged operators act on
$p$-dimensional closed submanifolds~\cite{NOc0605316, NOc0702377, GW14125148,
M220403045, CD220509545}. A $p$-symmetry is mathematically described by a
$(p+1)$-group\footnote{Here we will consider only pure ${p}$-symmetries, and
not the more general ${p}$-group symmetries where there are multiple symmetries
of different degrees that mix~\cite{BH180309336,
CI180204790,BCH221111764}.}~\cite{KT13094721, S150804770, ZW180809394}. Just
like global symmetries, higher-symmetries can be spontaneously
broken~\cite{L180207747}, can be anomalous~\cite{HT200311550}, and can be
gauged~\cite{HS181204716}.

Generic lattice Hamiltonians do not commute with closed string, membrane, \etc\ operators and thus do not have exact higher-symmetries. Instead, the lattice models with exact higher-symmetries are quite special. For instance, the Hamiltonians of many exactly soluble models~\cite{K032,W0303,LW0316,Y10074601,B11072707,WW1132} with topological orders~\cite{W8987,WN9077,W9039} have exact higher-symmetries. While higher-symmetries are typically not exact UV symmetries, they can nevertheless be \textit{emergent} symmetries occurring in the IR. Intuitively, this is because at high energies the charged $p$-dimensional closed objects can become open, and ${(p-1)}$-dimensional excitations living on their boundary explicitly breaks the $p$-symmetry. At energies smaller than the gap of the excitations that explicitly break a higher-symmetry, the corresponding higher-symmetry can emerge. While emergent 0-symmetries are typically approximate, emergent higher-form symmetries can exactly constraint the IR despite not being UV symmetries~\cite{FNN8035, HW0541, PS08010587, W181202517, SS200310466, IM210612610, COR220205866, HK221011492, VBV221101376,PW230105261}. In this sense, emergent higher-form symmetries are exact emergent symmetries~\cite{PW230105261}.

This gives rise to an interesting scenario where some low-energy excitations condense, while the higher-symmetry breaking excitations remain to have a large energy gap. If the condensed phase happens be a short-range entangled state~\cite{CGW1038} with the exact emergent higher-symmetry, it can be an SPT phase protected by the exact emergent higher symmetry~\cite{W181202517}. We denote such a higher SPT phase as an $n$-SPT phase if it is protected by an $n$-symmetry. 

In particular, suppose a higher-form symmetry emerges at ${E < E_{\text{mid-IR}}}$, is not spontaneously broken, and is realized anomalously on the boundary. A corresponding nontrivial SPT order could also emerge at ${E < E_{\text{mid-IR}}}$ and cause the system to be in an SPT phase. The nontrivial bulk SPT order allows an IR observer to turn on background gauge fields due to an anomaly inflow mechanism. The bulk theory for said IR observer would be an invertible topological field theory, called the SPT invariant, which characterizes the SPT order and its universal physical properties (i.e., through a generalized magneto-electric effect~\cite{MF220607725}). Furthermore, since emergent higher-form symmetries are exact emergent symmetries~\cite{PW230105261}, no local IR measurements could reveal that the high-form symmetry is not exact in the UV. 

However, according to a UV observer, the bulk theory would not follow the typical SPT lore since it is the IR degrees of freedom forming an SPT, not the UV degrees of freedom. It is conceivable that a UV observer could directly probe the topological response of the SPT by measuring UV degrees of freedom in a very particular way to couple to the IR degrees of freedom. Nevertheless, there are still direct signs of the emergent SPT order at the boundary, even in the UV. In the context of the SPTs we consider here, the boundary has nontrivial abelian topological order and thus this UV observer could measure the anyon excitations and their nontrivial braiding. The gap of the anyons would be on the scale ${E_{\text{mid-IR}}}$, and their presence would explicitly break the higher-form symmetries in the UV. However, their braiding would nevertheless reflect the 't Hooft anomaly structure and thus the emergent SPT order.

In this paper, we extend the work presented in \Rf{W181202517} and further investigate this mechanism for creating SPT phases protected by emergent higher-symmetries. In particular, we consider abelian gauge theory in 3+1D which at energies smaller than the gauge charge's and monopole's gap has two exact emergent $U(1)$ 1-symmetries (which we denote as $U(1)^{(1)}$) commonly denoted as the electric and magnetic symmetries. In the strong coupling limit, the gauge theory is in a short-range entangled gapped confined phase where the monopoles condense and the magnetic $U(1)^{(1)}$ symmetry is explicitly broken. However, at energies below the gauge charge gap, the electric symmetry is still present in the confined phase. This gives rise to the aforementioned scenario and the possibility that the confined phase has nontrivial 1-SPT order protected by the emergent electric symmetry.

In fact, here we will show that with $2\pi$-quantized topological terms, the
confined phase of abelian gauge theory has nontrivial
emergent 1-SPT orders. Usually, a topological term can affect the dynamics of
the strong coupling limit in a very non-trivial way, and can make it impossible
to calculate the physical properties (such as energy gap) in this limit.
However, a $2\pi$-quantized topological terms are much easier to
handle~\cite{CGL1204,OCX1226,X1321,BRX1315}, and we can still determine the
strong coupling limit to be a short-range entangled gapped confined phase.

The remaining of this paper is organized as follows. In section~\ref{notation}
we introduce the notations used in this paper. In section~\ref{Z2SPT}, we
present a simple example of a nontrivial 1-SPT order in the confined phase of
3+1D $\Z_2$ gauge theory. In doing so, we review the cochain lattice field
theory formalism and techniques which we use throughout the rest of the paper.
Then, in section~\ref{U1SPT} we consider the same scenario but in 3+1D abelian
gauge theory with $\kappa$-types of $U(1)$ gauge fields and $2\pi$-quantized
topological terms. Using the bosonic lattice model we develop, we find that the
total emergent electric symmetry ${U(1)^{(1)}\times U(1)^{(1)}\times\cdots}$
below the  gauge charges' gaps and the  monopoles' gaps is reduced to
${\Z_{k_1}^{(1)} \times \Z_{k_2}^{(1)} \times \cdots}$ at energies above the
monopoles' gaps (but still blow the gauge charges' gaps).  Subsequently, we
find that the confined phase of $U^\ka(1)$ gauge theory with $2\pi$-quantized
topological terms has nontrivial ${\Z_{k_1}^{(1)} \times \Z_{k_2}^{(1)} \times
\cdots}$ 1-SPT order. We construct the associated 1-SPT invariant and consider
some examples.

\section{Notations and conventions}
\label{notation}

In this paper, we will use the notion of cochain, cocycle, and coboundary, as well as their higher cup product $\hcup{k}$ and Steenrod square $\gSq^k$. A pedagogical introduction aimed at physicists of chains and cochains along with the cup product ${\smile\equiv\hcup{0}}$ and higher cup products $\hcup{k}$ can be found in the Appendix of
\Rf{TW190802613}. We will abbreviate the cup product $a\smile b$ as $ab$ by
dropping $\smile$. We will also use $\se{n}$ to mean equal up to a multiple of
$n$, and use $\se{\dd }$ to mean equal up to $\dd f$ (\ie up to a coboundary).  An important identity which we will repeatedly use is that for cochains $f_m, h_n$, 
\begin{align}
\label{cupkrel}
& \dd( f_m \hcup{k} h_n)=
\dd f_m \hcup{k} h_n +
(-)^{m} f_m \hcup{k} \dd h_n+
\\
& \ \ \
(-)^{m+n-k} f_m \hcup{k-1} h_n +
(-)^{mn+m+n} h_n \hcup{k-1} f_m .
\nonumber 
\end{align}

Furthermore, in this paper we will deal with many $\Z_n$-valued quantities. We will denote
them as, for example, $a^{\Z_n}$. However, we will always lift the $\Z_n$-value
to $\Z$-value, so the value of $a^{\Z_n}$ has a range from $ -\toZ{\frac n2}$
to $ \toZ{\frac n2}$, where  $\toZ{x}$ denotes the integer that is closest to
$x$ (if two integers have the same distance to $x$, we will choose the smaller
one, \eg $\toZ{\frac12}=0$).  In this case, the expression like
$a^{\Z_n}a^{\Z_m}$ makes sense.

\section{$\Z_2^{(1)}$ 1-SPT order in 3+1D theories}
\label{Z2SPT}

In this section, we review one of the simplest ways to realize $\Z_2^{(1)}$ 1-SPT order in 3+1D~\cite{ZW180809394,W181202517}. Our purpose of doing so is to introduce the formalism we use and warm-up in a simple context before beginning section~\ref{U1SPT}, where things get more involved. We start by considering a twisted $\Z_2$ 2-gauge theory. By considering its confined phase, we then construct a model with $\Z_2^{(1)}$ ${\text{1-SPT}}$ order by ``ungauging'' the twisted $\Z_2$ 2-gauge theory. The $\Z_2^{(1)}$ 1-SPT order is exact in this model, but survives elsewhere in the confined phase diagram as an exact emergent $\Z_2^{(1)}$ 1-SPT, existing at energies much smaller than the $\Z_2$ gauge charge gap.

\subsection{Twisted $\Z_2$ 2-gauge theory}

To construct a 3+1D bosonic model that realizes $\Z_2^{(1)}$ 1-SPT order, we first construct a local bosonic model with topological order
described by a $\Z_2$ ${\text{2-gauge}}$ theory. We triangulate spacetime $\cM^4$ and, working in the Euclidean signature, consider the lattice path integral of cochain fields~\cite{W161201418}. The bosonic degrees of freedom for the $\Z_2$ ${\text{2-gauge}}$ theory are described by a $\Z_2$-valued 2-cochain field $b^{\Z_2}$. As a 2-cochain, $b^{\Z_2}$ is a map from 2-chains to $\Z_2$, as opposed to ${\text{1-gauge}}$ theory which is described by a 1-cochain field.

Consider the local bosonic model:
\begin{equation}
Z(\cM^4, g) = \sum_{b^{\Z_2}} \ee^{-\frac 1{2g} \sum_{ijkl}  
\left( \dd b^{\Z_2}_{ijkl}  - 2\toZ{\frac12 \dd b^{\Z_2}_{ijkl} }\right)},
\end{equation}
where ${\sum_{ijkl}}$ sums over all spacetime 3-simplices for a fixed triangulation, and $\sum_{b^{\Z_2}}$ sums
over all the 2-cochain field corresponding to the path integral. In the exactly solvable limit ${g\to 0}$, the path integral becomes
\begin{equation}\label{ZbZ2}
 Z = \sum_{\dd b^{\Z_2}\se{2}0} 1,
\end{equation}
where ${\dd b^{\Z_2}\se{2}0}$ means ${\dd b^{\Z_2}=0~\mod~2}$. Now, $Z$ captures the topological order described by the deconfined phase of pure $\Z_2$
2-gauge theory. However, we note that in ${3+1}$D, $\Z_2$ 2-gauge theory is dual to $\Z_2$ 1-gauge theory\footnote{In $\Z_2$ 2-gauge theory in 3+1D, loop excitations carry $\Z_2$ gauge charge while particle excitations carry the $\Z_2$ gauge flux. On the other hand, in $\Z_2$ 1-gauge theory in 3+1D, particles carry the $\Z_2$ gauge charge and loops carrying the $\Z_2$ gauge flux. The duality between $\Z_2$ 2-gauge theory and $\Z_2$ 1-gauge theory in ${3+1}$D simply switches which excitations are called gauge charges and which are called gauge fluxes.}. Thus, the topological order is also described by $\Z_2$ 1-gauge theory, which is $\Z_2$ topological order. In fact, Eq.~\eqref{ZbZ2} is the ${3+1}$D toric code.

We now consider the equivalent limit in a twisted $\Z_2$ 2-gauge theory. This is described by a different bosonic model, which is Eq.~\eqref{ZbZ2}  but with the 1 replaced with the action amplitude $\ee^{\ii \pi \int_{\cM^4} (b^{\Z_2})^2}$. Indeed, the path integral is
\begin{align}
\label{ZbZ2t}
 Z(\cM^4) &= \sum_{\dd b^{\Z_2}\se{2}0} \ee^{\ii \pi \int_{\cM^4} (b^{\Z_2})^2 },
\end{align}
where we use the shorthand ${(b^{\Z_2})^2 \equiv   b^{\Z_2}\smile b^{\Z_2}}$ and $\int_{\cM^4}$ is a sum over all 4-simplicies of $\cM^4$. Note that this action amplitude is correctly invariant under the gauge transformation ${b^{\Z_2} \to b^{\Z_2} + 2n^{\Z}}$, where $n^{\Z}$ is an arbitrary $\Z$-valued 2-cochain. The twisted $\Z_2$ 2-gauge theory realizes a twisted $\Z_2$ topological order where the $\Z_2$ charges are fermions~\cite{W161201418}.

\subsection{Lattice model with $\Z_2$ 1-SPT order}

We now use the twisted $\Z_2$ 2-gauge theory in Eq.~\eqref{ZbZ2t} to obtain a
local bosonic model that realizes a 1-SPT order protected by the $\Z_2$
2-group.  The classifying space of the $\Z_2$ 2-group is a topological space
denoted by $B(\Z_2,2)$, which satisfies ${\pi_2(B(\Z_2,2))=\Z_2}$ and
${\pi_n(B(\Z_2,2))=0}$ for $n\neq 2$.  The associated symmetry is a $\Z_2$
1-symmetry, which we denote as $\Z_2^{(1)}$.

The $\Z_2$ 2-gauge theory can be ``ungauged'' by first parameterizing the
dynamical 2-cochain field $b^{\Z_2}$ as
\begin{equation}
 b^{\Z_2} = B^{\Z_2} + \dd a^{\Z_2},
\end{equation}
where $a^{\Z_2}$ is a $\Z_2$-valued 1-cochain field describing the pure 2-gauge fluctuations. However, we now reinterpret the meaning of $B^{\Z_2}$ and $a^{\Z_2}$ by treating $a^{\Z_2}$ as the dynamical field and $B^{\Z_2}$ as a $\Z_2$-valued 2-cocycle background field. This produces a new local bosonic model whose resulting path integral is obtained from the twisted $\Z_2$ 2-gauge theory Eq.~\eqref{ZbZ2t}:
\begin{equation}
\label{Bda}
 Z (\cM^4, B^{\Z_2}) = \sum_{a^{\Z_2}} 
\ee^{\ii \pi \int_{\cM^4} (B^{\Z_2}+\dd a^{\Z_2})^2}.
\end{equation}
This path integral is invariant under the gauge transformation
\begin{align*}
 a^{\Z_2} &\to a^{\Z_2} +\al^{\Z_2},\\
 B^{\Z_2} &\to B^{\Z_2} - \dd\al^{\Z_2}.
\end{align*}
$B^{\Z_2}$ describes the symmetry-twist of the $\Z_2^{(1)}$ 1-symmetry. Turning off the background symmetry-twist field, and hence ungauging the $\Z_2^{(1)}$ 1-symmetry, the model becomes
\begin{align}
\label{dada}
 Z (\cM^4,0) &= \sum_{a^{\Z_2}} 
\ee^{\ii \pi \int_{\cM^4} (\dd a^{\Z_2})^2  },
\end{align}
which has an exact $\Z_2^{(1)}$ 1-symmetry is generated by $\Z_2$-valued 1-cocycles
$\al^{\Z_2}$:
\begin{align}
\label{1symmZ2}
 a^{\Z_2} \to a^{\Z_2} +\al^{\Z_2}, \ \ \ \
\dd \al^{\Z_2} \se{2} 0.
\end{align}
By construction, there is no obstruction to gauging the $\Z_2^{(1)}$ 1-symmetry and therefore the
$\Z_2^{(1)}$ 1-symmetry is anomaly-free. This can further be seen from the fact that the path integral $Z (\cM^4)$ is invariant under the $\Z_2^{(1)}$
transformation even when $\cM^4$ has boundary. 

Using that ${\int_{\cM^4} (\dd a^{\Z_2})^2 = \int_{\p\cM^4} a^{\Z_2}\dd a^{\Z_2}}$, when spacetime $\cM^4$ is closed (i.e., ${\p\cM^4 = \emptyset}$) then ${\int_{\cM^4} (\dd a^{\Z_2})^2 = 0}$. Therefore, the action amplitude
${\ee^{\ii \pi \int_{\cM^4} (\dd a^{\Z_2})^2}=1}$ and so for a closed spacetime
\begin{equation}\label{volumeZ}
 Z (\cM^4,0) = \sum_{a^{\Z_2}} 
\ee^{\ii \pi \int_{\cM^4} (\dd a^{\Z_2})^2  } = 2^{N_e},
\end{equation}
where $N_e$ is the number of the edges in the spacetime complex $\cM^4$.
According to a conjecture presented in \Rf{KW1458}, this implies that the
ground state of the model Eq.~\eqref{dada} has no topological order (\ie is
short range entangled).

Since the ground state has $\Z_2^{(1)}$ 1-symmetry and no topological order, it
may have a $\Z_2^{(1)}$ 1-SPT order, which are classified by the fourth
cohomology group ${H^4(B(\Z_2,2),\RZ) =
\Z_4}$~\cite{ZW180809394,W181202517,WW181211967,TW190802613}.  To see which
1-SPT order is realized, we note that the SPT order is characterized by the
volume-independent partition function
\begin{equation}
Z^{\text{top}}(\cM^4,B^{\Z_2}) 
=  \dfrac{Z (\cM^4,B^{\Z_2})}{Z (\cM^4,0)},
\end{equation}
which is called the SPT-invariant~\cite{K1459, K1467, W1447, HW1339}.  We
compute the 1-SPT invariant from Eq. \eq{Bda}, by integrating out $a^{\Z_2}$ for
closed spacetime $\cM^4$ and for $\dd B^{\Z_2}=0$ mod 2.  Using
Eq.~\eqref{volumeZ} and the fact that $B^{\Z_2}$ is a cocycle, we find that
\begin{align}\label{Z2SPTinv}
Z^{\text{top}}(\cM^4,B^{\Z_2}) 
&=\ee^{\ii \pi \int_{\cM^4} (B^{\Z_2})^2  }
\nonumber\\
&=\ee^{\ii  \frac{m}{4} 2\pi\int_{\cM^4} \gSq^2(B^{\Z_2})  }\Big|_{m=2}.
\end{align}
Here, the generalized Steenrod square $\gSq^{k}$ is defined as 
\begin{align}
\label{Sqdef}
 \gSq^{k}(c_l) \equiv c_l\hcup{l-k} c_l +  c_l\hcup{l-k+1} \dd c_l,
\end{align}
where $c_l$ is any $l$-cochain. 
From the above 1-SPT invariant, we see that the model defined by Eq.~\eqref{dada}
realizes a $\Z_2^{(1)}$ 1-SPT order that corresponds to $2\in \Z_4$ in the
confined phase.

\subsection{Emergent $\Z_2^{(1)}$ 1-SPT order in the confined phase of $\Z_2$
gauge theory}

\label{emergentZ21SPT}

The fact that the theory Eq.~\eqref{dada} has an exact $\Z_2^{(1)}$ symmetry makes it rather special. Indeed, for a typical condensed matter model, the lattice theory would be more like
\begin{align}
\label{model}
 Z[\cM^4,g,h]  &= \sum_{a^{\Z_2}} 
\ee^{\ii \pi \int_{\cM^4} (\dd a^{\Z_2})^2  -h\sum_{ij} (a^{\Z_2})_{ij}} \times
\\
&\ \ \ \ \ \ \ \
\ee^{-\frac 1{2g} \sum_{ijk} 
( \dd a^{\Z_2})_{ijk} -2\toZ{\frac12 (\dd a^{\Z_2})_{ijk} }
 } 
,
\nonumber 
\end{align}
where $\sum_{ijk}$ sums over all the triangles and
$\sum_{ij}$ sums over all the edges of $\cM^4$. This path integral does not have the $\Z_2^{(1)}$ symmetry \eqref{1symmZ2}, it is explicitly broken by the $h$ term. Only when ${h=0}$ does Eq.~\eqref{model} have the $\Z_2^{(1)}$
symmetry. Thus, at first glance, when ${h\neq 0}$ this generic model does not realize a $\Z_2^{(1)}$ SPT state since it does not even have the symmetry. However, while the $\Z_2^{(1)}$ symmetry is no longer a UV symmetry, for small $|h|$ the low-energy sectors of the Hilbert space enjoys an \textit{exact emergent} $\Z_2^{(1)}$ symmetry. Indeed, since only the motion of the $\Z_2$ charge excitations can break the $\Z_2^{(1)}$ 1-symmetry (i.e., the $h$ term), a $\Z_2^{(1)}$ symmetry emerges at energies much smaller than the $\Z_2$ charge excitation gap.

\begin{figure}[t]
\begin{center}
\includegraphics[width=.48\textwidth]{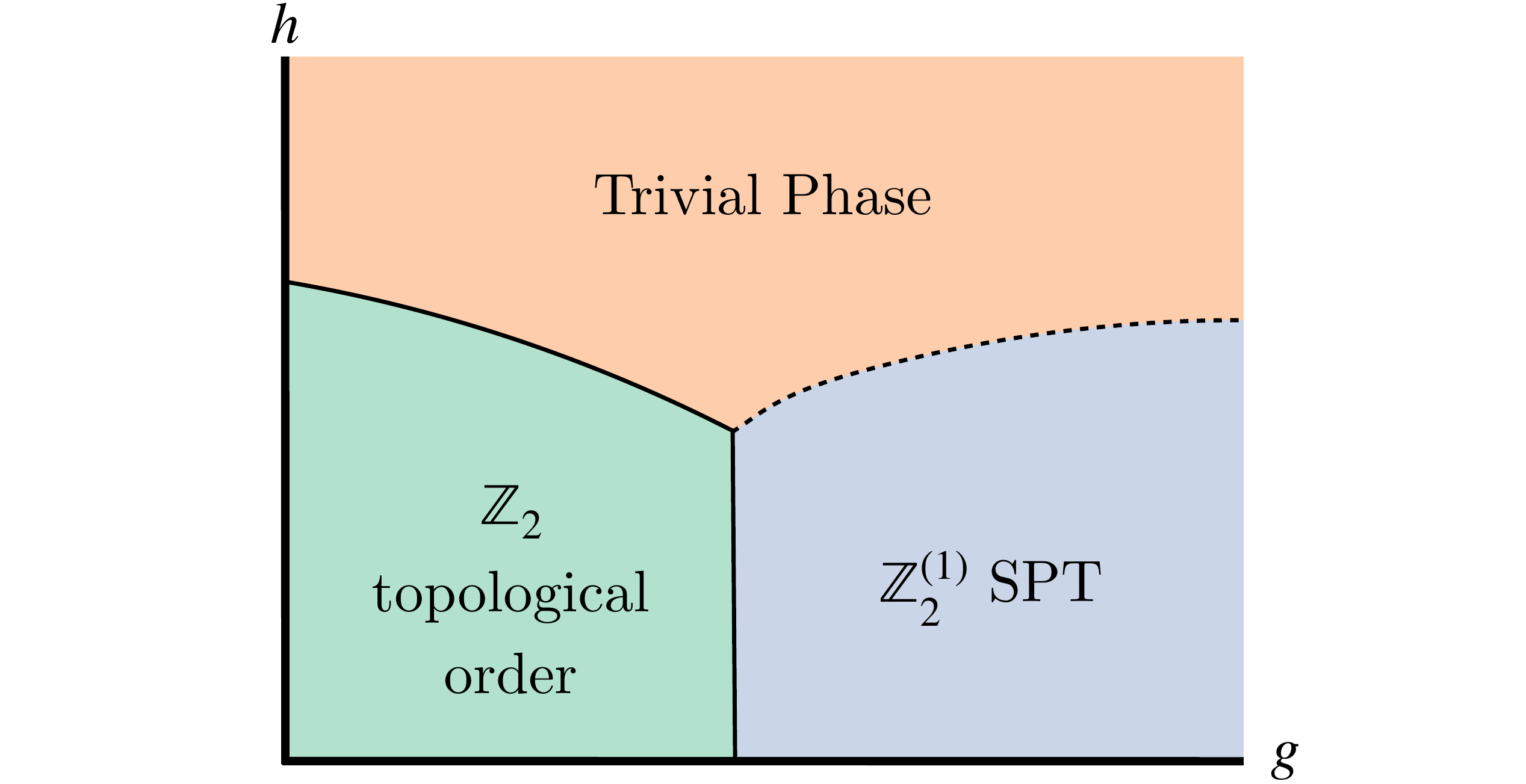}
\end{center}
\caption{The schematic phase diagram of the model described by Eq.~\eqref{model}. There
is a topologically ordered phase described by the deconfined phase of $\Z_2$
gauge theory (shown in green), and a gapped short-range entangled phase (corresponding to the regions shaded in orange and purple). At energies below the $\Z_2$ gauge charge gap, the confined phase (colored in purple) has an exact emergent $\Z_2^{(1)}$ symmetry, while the symmetry is absent at all energy scales in the trivial phase (colored in orange). At ${h=0}$, this $\Z_2^{(1)}$ symmetry in the confined phase is exact. Due to the $2\pi$-quantized topological term, the confined phase has a nontrivial SPT order protected by this exact emergent $\Z_2^{(1)}$
symmetry. The SPT invariant describing this 1-SPT order is given by
Eq.~\eqref{Z2SPTinv}} \label{Z2ph} 
\end{figure}

When $|h|\ll 1$ and $|g|\ll 1$, the model Eq.~\eqref{model} realizes the $\Z_2$
topological order described by the deconfined phase of $\Z_2$ gauge theory.  As
we increase $g$, it will undergo a phase transition into a confined phase with
short-range entanglement.  Let us assume this transition is continuous (if it
is not, we can modify the model to make the confinement transition nearly
continuous). Then, approaching the transition, the $\Z_2$-flux fluctuations are
low energy excitations while the $\Z_2$ charge excitations remain to have a
large energy gap. This is exactly the scenario for the exact emergent
$\Z_2^{(1)}$ symmetry. So in the confined phase (\ie when
the $\Z_2$ charge excitations energy gap remains large), the model realizes an
1-SPT order protected by the exact emergent $\Z_2^{(1)}$ 1-symmetry, described
by $2\in\Z_4$.

The low-energy effective theory describing the phase with emergent 1-SPT order is Eq.~\eqref{dada}. Introducing the Poincar\'e dual\footnote{The Poincar\'e dual of the $(D-p)$-cycle $C$ with respect to the $D$-dimensional complex $M$, denoted as $\hat{C}$, satisfies ${\int_C a = \int_M a~\hat{C}}$, where ${a}$ is any ${(D-p)}$-cochain.} of $\dd a^{\Z_2}$, denoted as $\hat{f}$, the lattice action ${\int_{\cM^4} (\dd a^{\Z_2})^2 }$ is equal to the intersection number of $\hat{f}$: ${\int_{\cM^4} (\dd a^{\Z_2})^2 = \#(\hat{f}\cdot\hat{f})}$. We note that $\hat{f}$ corresponds to the world sheets of $\Z_2$ flux loops so ${\#(\hat{f}\cdot\hat{f})}$ is the intersection number of $\Z_2$ flux world sheets in spacetime. The topological term ${\ee^{\ii \pi \int_{\cM^4}(\dd a^{\Z_2})^2 }}$ is therefore 
\begin{align}
 \ee^{\ii \pi \int_{\cM^4} (\dd a^{\Z_2})^2 }
=(-1)^{\#(\hat{f}\cdot\hat{f})} .
\end{align}

In general, the path integral of a $\Z_2$ gauge theory may or may not contain the topological term ${(-1)^{\#(\hat{f}\cdot\hat{f})}}$. When the topological term is included, then the confined phase of the $\Z_2$ gauge theory will be a 1-SPT state protected by the exact emergent
$\Z_2^{(1)}$ 1-symmetry. However, when the path integral does not include the
topological term, then the confined phase of
the $\Z_2$ gauge theory will be a trivial SPT
state of the exact emergent $\Z_2^{(1)}$ 1-symmetry. Therefore, in a 3+1D
$\Z_2$ gauge theory, by adjusting the presence or the absence of the
topological term ${(-1)^{\#(\hat{f}\cdot\hat{f})}}$ (\ie the intersection term for the $\Z_2$-flux world sheet), we can control the presence
or the absence of the 1-SPT order protected by the exact emergent $\Z_2^{(1)}$
1-symmetry in the confined phase.

\subsection{Using confined phases of 3+1D $\Z_n$ gauge theory to realize
$\Z_n^{(1)}$ 1-SPT orders for even $n$}

For simplicity, we've presented the above in the $\Z_2$ case, but it can easily be generalized by replacing $\Z_2$ with $\Z_n$, where $n$ is a positive even integer. Indeed, introducing the $\Z_n$-valued 1-cochain field $a^{\Z_n}$, the generalized generic lattice model is
\begin{align}
\label{Znmodel}
 Z  &= \sum_{a^{\Z_n}} 
\ee^{\ii \pi \frac{m}{n} \int_{\cM^4} \gSq^2(\dd a^{\Z_n})   -h\sum_{ij} (a^{\Z_n})_{ij}}\times
\\
&\ \ \ \ \ \ \ \
\ee^{-\frac 1{ng} \sum_{ijk} 
( \dd a^{\Z_n})_{ijk} -n\toZ{\frac1n (\dd a^{\Z_n})_{ijk} }
 } 
,
\nonumber 
\end{align}
where the topological term is now proportional to ${\int_{\cM^4} \gSq^2(\dd
a^{\Z_n})}$.  In \Rf{TW190802613}, it was shown that ${\gSq^2(\dd a^{\Z_n} +nb)
\se{2n} \gSq^2(\dd a^{\Z_n})}$ and that ${\ee^{\ii \pi \frac{m}{n} \int_{\cM^4}
\gSq^2(\dd a^{\Z_n})  } = 1}$ when $\cM^4$ is closed. Thus, the inclusion of
the topological term ${\ee^{\ii \pi \frac{m}{n} \int_{\cM^4} \gSq^2(\dd
a^{\Z_n})  }}$ does not affect the local dynamics of the model. 

As a result, when $|h|\ll 1$ and $|g|\ll 1$, the model \eqref{Znmodel} realizes
the $\Z_n$ topological order described by the deconfined phase of a $\Z_n$
gauge theory.  As we increase $g$, the model will undergo a confinement phase
transition. Assuming the transition is continuous, near the transition the
$\Z_n$-flux energy gap is much smaller than the gap for the $\Z_n$ charge
excitations. So near the transition, the model has an exact emergent $\Z_n^{(1)}$
1-symmetry, at energies much less than  $\Z_n$ charge energy gap. After the
confinement transition, the confined phase has 1-SPT order protected by the
exact emergent $\Z_n^{(1)}$ 1-symmetry and described by $m\in
H^4(B(\Z_n,2);\RZ)=\Z_{2n}$ for even $n$\footnote{The lattice model
\eqref{Znmodel} is well defined even for $m\notin \Z$.  However, when $m\notin
\Z$ it is not clear if the model has a gapped confined phase when $|g|\gg 1$.}.
Here $B(\Z_n,2)$ is the classifying space of the $\Z_n$ 2-group describing
$\Z_n^{(1)}$ 1-symmetry.  It satisfies $\pi_2(B(\Z_n,2))=\Z_n$ and
$\pi_i(B(\Z_2,2))=0,\ i\neq 2$.  The physical consequence of  $\Z_n^{(1)}$
1-SPT order, such as boundary states, as well as a Hamiltonian description of this phase, was discussed in \Rf{TW190802613}.

\section{Emergent ${\Z_{k_1}^{(1)}\times
\Z_{k_2}^{(1)}\times ... }$ 1-SPT order in a 3+1D $U^\ka(1)$ bosonic model}\label{U1SPT}

In last section, we saw how $\Z_{2n}^{(1)}$ 1-SPT state can be realized in the
confined phase of 3+1D $\Z_n$ gauge theory. Now
we investigate more complicated 1-SPT orders which are protected by finite
1-symmetries. In this section, we will construct a 3+1D bosonic model, that
corresponds to lattice $U^\ka(1)$ ``gauge theory'' with a $2\pi$-quantized
topological term.  We will show that, due to the topological term, the model
has a reduced $ \Z_{k_1}^{(1)} \times \Z_{k_2}^{(1)} \times \cdots $
1-symmetry.  We will also show that the confined phase of the $U^\ka(1)$ gauge
theory can have a 1-SPT orders protected by the $\Z_{k_1}^{(1)} \times
\Z_{k_2}^{(1)} \times \cdots $ 1-symmetry.

\subsection{3+1D $U^\ka(1)$ pure gauge field theory and $2\pi$-quantized topological term}

Before we consider the bosonic lattice model, we first consider the corresponding continuum theory. We do so in a timely, but non-rigorous, fashion to see how the results from the lattice theory which we present in the next sections are hinted towards in the continuum theory. It will set the stage for the lattice theory where the formal manipulations are much more involved than those in the field theory.

We consider the theory described by the Euclidean action
\begin{equation}\label{eqn:actionWithTheta}
S = \frac{1}{2g^2}\sum_{I}\int_{M^4} f^{I} \wdg \hstar f^{I} + S_\text{top},
\end{equation}
where $a^{I}$ with ${I = 1,\ldots, \kappa}$ are $\RZ$-valued 1-form
fields\footnote{Typically the $U(1)$ connection is a map ${A:\R^4\mapsto
\R/2\pi\Z}$. We define $a = A/2\pi$ to absorb factors of $2\pi$ and match the
convention used in the bosonic lattice model. In terms of the coupling constant
$g$, the typical $U(1)$ coupling constant is $e = 2\pi g$.}, the 2-form
curvature $f^{I} = \dd a^{I}$, and ${k_{IJ}\in\Z}$. The first term is Maxwell's
kinetic term and the second term is the $2\pi$-quantized topological term
(topological in the sense that it is independent of the metric)
\begin{equation}\label{contTopTerm1}
S_{\text{top}} = -2\pi\ii  \sum_{I\leq J}k_{IJ}\int_{M^4} f^I \wdg f^J,
\end{equation}
Furthermore, the quantity ${k_{IJ}\int f^I\wdg f^J}$ is quantized as an integer
when $M^4$ is closed. Thus, for closed $M^4$ the action amplitude of the
topological term is unity, but for $M^4$ with a boundary it can have a
nontrivial effect.

Since the action depends only on $f^{I}$, it is left unchanged by
\begin{equation}
\label{a1Sym}
a^{I} \to a^{I} + \Gamma^{I}, \ \ \ \ \dd\Gamma^{I} = 0.
\end{equation}
This corresponds to a real symmetry transformation (not a gauge transformation)
when ${\oint \Gamma^{I} \neq \Z}$. 
Since
there are $\kappa$ fields $a^{I}$ (i.e., ${I = 1,\cdots,\ka}$),
Eq.~\eqref{a1Sym} is associated with $\kappa$ different $U(1)^{(1)}$ 1-form
symmetries: ${U(1)^{(1)}\times U(1)^{(1)}\times\cdots}$.
The associated Noether current can be found
by introducing a background symmetry twisted field $\cB^I$ in Lorentzian signature
and having ${\dd a^I \to \dd a^I - \cB^I}$. Noting that the conserved current
$J^I$ minimally couples to $\cB^I$ as ${\int \cB^I ~\wdg \hstar J^I}$, we find
that for the transformation of the $I$th field:
\begin{equation}
J^{I} =  \dfrac{1}{g^2}f^{I} + 2 \pi \sum_J K_{IJ}\hstar f^{J},
\end{equation}
where $K_{IJ}$ is given by
\begin{align}
 K_{II}=2k_{II}, \ \ \ \ \ \
K_{IJ}=K_{JI}=k_{IJ}, \ \  I < J.
\end{align}
The fact that the current is conserved means that ${\dd^\dagger J^I = 0}$, where
${\dd^\dagger = \hstar\dd\hstar}$ in the adjoint of $\dd$. 

In the above analysis of the symmetry, we consider the field theory
without $U(1)$ charges and $U(1)$ monopoles.  This ${U(1)^{(1)}\times
U(1)^{(1)}\times\cdots}$ is really an exact emergent 1-form symmetry at energies below the
$U(1)$ charge gaps and the monopole gaps. Indeed, at energies above the $U(1)$ gauge charge gaps, terms like ${\int a^I \wdg \hstar j^I}$ will contribute to the action and this symmetry will be explicitly broken.

Let's now introduce the 1-form ${j^{I}_m = \hstar\dd f^{I}}$, the Dirac
monopole current density associated with the $I$th field $a^{I}$and the Poincar\'e dual of ${\hstar j^{I}_m}$ gives
the world-line of the monopole. The continuity equation ${\dd^\dagger J^{I} =
0}$ then implies that
\begin{equation}\label{GWlaw}
\dfrac{1}{g^2}\dd^\dagger f^{I} = 2\pi K_{IJ}j_m^{J}.
\end{equation}
The effect of the nonzero righthand side is a generalized version of the Witten
effect where $U(1)$ monopoles of the $J$th field carries $K_{IJ}$ units of the
$I$th $U(1)$ gauge charge.

The presence of magnetic monopoles complicates things.  At energies below the
$U(1)$ charge gaps but above the monopole gaps, due to the topological term,
the monopoles fluctuations imply $U(1)$ charge fluctuations.  This may break
the ${U(1)^{(1)}\times U(1)^{(1)}\times\cdots}$ 1-form symmetry to a smaller
symmetry.

In the continuum, monopole configurations can be easily considered by
parametrizing the curvature as ${f^I = \dd \tilde{a}^I + G^I}$. The 1-form
fields $\tilde{a}^I$ describe the smooth local fluctuations of $a^I$ and
satisfy the Bianchi identity $\dd (\dd \tilde{a}^I) = 0$, while the 2-form
fields $G^I$ capture the singular monopole configurations and satisfy ${j^I_m =
\star\dd G^I}$. At energies above the monopole gap, the field theory that
describes the lattice model instead has the topological term
\begin{equation}\label{contTopTerm2}
\begin{aligned}
S_\text{top} &= -2\pi\ii \sum_{I\leq J}   k_{IJ} \int_{M^4}   \dd\tilde{a}^I \wdg\dd\tilde{a}^J  +
G^I \wdg G^J \\
&\hspace{70pt}  -2\pi \ii   \sum_{I, J} K_{IJ} 
\int_{M^4}
\tilde{a}^I \wdg \hstar j_m^J.
\end{aligned}
\end{equation}
This is equivalent to Eq.~\eqref{contTopTerm1} up to a boundary term. For all
practical purposes, we may treat the density in Eq.~\eqref{contTopTerm2} as the
definition of $f^I\wdg f^J$ for energies above the monopole gap. This
distinction is important as the ${U(1)^{(1)}\times U(1)^{(1)}\times\cdots}$
symmetry of Eq.~\eqref{contTopTerm1} is broken down to a finite subgroup in
Eq.~\eqref{contTopTerm2}, agreeing with the symmetries of the lattice model we
study.

Indeed, above the monopole energy gap, Eq.~\eqref{contTopTerm2} is invariant under the transformation
\begin{equation}
\tilde{a}^{I} \to \tilde{a}^{I} + \Gamma^{I}, \ \ \ \ \sum_{I} K_{IJ} \oint_{C^1}\Gamma^I \in \Z,
\ \ \ \ \dd\Gamma^{I} = 0,
\end{equation}
for any closed 1-submanifold $C^1$. The additional restriction ${\sum_{I} K_{IJ} \oint_{C^1}\Gamma^I \in \Z}$ ensures that the action amplitude ${\ee^{2\pi \ii   \sum_{I, J} K_{IJ} 
\int_{M^4}
\tilde{a}^I \wdg \hstar j_m^J}}$ is invariant since ${\oint \hstar j^I_m \in \Z}$. We note that this term in Eq.~\eqref{contTopTerm2} also recovers the Gauss-Witten law Eq.~\eqref{GWlaw}. Thus, at a fixed point in spacetime, the values of allowed $\Gamma^I$ form a rational lattice $K^{-1}$. So, above the monopole gap the theory has the 1-symmetries ${\Z_{k_1}^{(1)}\times \Z_{k_2}^{(1)}\times \cdots}$, where $k_i$ are the diagonal elements of the Smith normal form for $K$. Below the monopole gap when $j^I_m$ vanishes, this constraint on $\Gamma^I$ does not apply so there is instead the aforementioned ${U(1)^{(1)}\times U(1)^{(1)}\times\cdots}$ symmetries.

Let's now turn on 2-form background fields $\cB^I$ that are the flat connections describing the twist of the ${\Z_{k_1}^{(1)}\times \Z_{k_2}^{(1)}\times \cdots}$ symmetry and satisfy the quantization conditions
\begin{equation}
\sum_{I}K_{IJ}  \oint_{C^2} \cB^I \in \Z,
\end{equation}
for any closed 2-submanifold $C^2$. We'll work locally at the level of differential forms, ignoring topological subtitles and monopoles. The background fields minimally couple to the dynamical fields $a^I$ by replacing the curvature $\dd a^I$ in the Euclidean action by ${\dd a^I - \cB^I}$. Making this replacement and taking the ${g\to\infty}$ limit, the action becomes
\begin{equation}\label{contTopActBCKG}
S = -2\pi\ii  \sum_{I\leq J}k_{IJ}\int_{M^4} (\dd a^I- \cB^I) \wdg (\dd a^J- \cB^J).
\end{equation}

We can use Eq.~\eqref{contTopActBCKG} to find the continuum SPT invariant which describes the 1-SPT order in the confined phase. Indeed, let's consider spacetime $M^4$ to be closed. Then, since we ignore monopoles and because ${\dd \cB_I = 0}$, integrating by parts we can rewrite the Euclidean action as
\begin{equation}
\begin{aligned}
S &=  -2\pi\ii  \sum_{I\leq J}k_{IJ}\int_{M^4} \cB^I\wdg \cB^J,\\
&=  -\ii\pi  \sum_{I, J}K_{IJ}\int_{M^4} \cB^I\wdg \cB^J.
\end{aligned}
\end{equation}
Thus, the path integral $Z$ in this limit is
\begin{equation}\label{PIcont1}
\begin{aligned}
Z[M^4,B^I] &= \int D[a^I]~e^{\ii\pi  \sum_{I,J}K_{IJ}\int_{M^4} \cB^I\wdg~\cB^J},\\
&=  \operatorname{Vol}^\kappa(\RZ)~e^{\ii\pi  \sum_{I,J}K_{IJ}\int_{M^4} \cB^I\wdg~\cB^J},
\end{aligned}
\end{equation}
where we've used that the action amplitude does not depend on the dynamical fields $a^I$ and introduced
\begin{equation*}
\operatorname{Vol}^\kappa(\RZ) = \int D[a^I].
\end{equation*}

The SPT invariant is given by the volume-independent part of the path integral 
\begin{equation}
Z^{\text{top}}(M^4,B^I)
=  \dfrac{Z(M^4,B^I)}{Z (M^4,0)}.
\end{equation}
Therefore, using Eq.~\eqref{PIcont1} we find that in the continuum theory the 1-SPT invariant is
\begin{equation}\label{contSPTinv}
Z^{\text{top}}(M^4,B^I) = e^{\ii\pi  \sum_{I,J}K_{IJ}\int_{M^4} \cB^I\wdg~\cB^J}.
\end{equation}
Thus, without much work we can characterize the 1-SPT order. However, in doing
so we ignored nontrivial fibre bundles and magnetic monopoles. In the remainder
of this section, we'll regulate this continuum theory by considering a bosonic
lattice model whose IR properties are described by the field theory. Using this
lattice model, we'll be able to recalculate the SPT invariant more rigorously
(see Eq.~\eqref{1SPTinv0}), and find lattice-dependent terms in addition to one
which captures Eq.~\eqref{contSPTinv}.

\subsection{Lattice Regularization of $U^\ka(1)$ gauge theory with $2\pi$-quantized topological
term}\label{latticemodel}

We now regulate the field theory discussed in the previous section by triangulating spacetime. The 1-form fields $a^I$ will be represented by $\RZ$-valued $1$-cochains $a_I^\RZ$. There are three key properties that the $U^\ka(1)$ gauge theory on a lattice must include:
\begin{enumerate}
\item Letting $m_I^\Z$ be an arbitrary $Z$-valued 1-cochain, the action amplitude is invariant under ${a^\RZ_I \to a^\RZ_I+m_I^\Z}$,
even when spacetime $\cM^4$ has a boundary;
\item When $\cM^4$ is closed, the action amplitude of the $2\pi$-quantized topological term becomes unity;
\item In the smooth field limit (the low energy limit) when ${\dd a^{\RZ}_J\sim \toZ{\dd a^{\RZ}_J}}$, which implies no
monopoles), the action amplitude reduces to its continuum limit $ \ee^{\ii 2\pi \int_{M^4} \sum_{I\leq J} k_{IJ}
f^I \wdg f^J }$.
\end{enumerate} 

Regularizing the Maxwell term on the lattice is straight forward, but the $2\pi$-quantized topological term is
highly non-trivial. Noting the relationship between the topological term and Chern-Simons theory in the continuum, this motivates us to define the $2\pi$-quantized topological term on the lattice as the derivative of the lattice Chern-Simons action. Indeed, we start with 2+1D $U^\ka(1)$
Chern-Simons theory on spacetime lattice $\cB^3$ obtained in \Rf{DW190608270}
\begin{equation}\label{CSlatt1}
\begin{aligned}
&
\hspace{-5pt}Z_{\text{CS}}=\int D[a^\RZ_I]\
\ee^{\ii 2\pi \hspace{-2pt}\sum\limits_{I\leq J} \hspace{-2pt}k_{IJ} \hspace{-4pt}\int\limits_{\cB^3} \hspace{-2pt} \dd \big(a^{\R/\Z}_I(a^{\R/\Z}_J-\toZ{a^{\R/\Z}_J})\big) }\\
&
\hspace{5pt}\times\hspace{-3pt}\ee^{\ii 2 \pi  \hspace{-2pt}\sum\limits_{I\leq J} \hspace{-2pt}k_{IJ} \hspace{-4pt}\int\limits_{\cB^3} \hspace{-2pt}
a^\RZ_{I} (\dd a^\RZ_{J} -\toZ{\dd a^\RZ_J})-\toZ{\dd a^\RZ_I}a^\RZ_J
}
\\
&
\hspace{5pt}\times\hspace{-3pt}\ee^{-\ii 2\pi \hspace{-2pt}\sum\limits_{I\leq J} \hspace{-2pt}k_{IJ} \hspace{-4pt}\int\limits_{\cB^3} \hspace{-2pt}  a^{\RZ}_J\hcup{1}\dd \toZ{\dd a^{\RZ}_I}}
\hspace{-2pt}\ee^{- \hspace{-2pt}\sum\limits_{I}\hspace{-2pt}\int\limits_{\cB^3} \hspace{-2pt} \frac{|\dd a^{\R/\Z}_I - \toZ{\dd a^{\R/\Z}_I}|^2}{g_3}}\hspace{-4pt},
\end{aligned}
\end{equation}
where $a^\RZ_I$ are the aforementioned $\RZ$-valued 1-cochain, the path-integral notation is shorthand for ${\int D[a^\RZ_I]= \prod_{ij,I}
\int_{-\frac12}^{\frac 12}\dd (a^\RZ_I)_{ij}}$, and $k_{IJ}\in\Z$. This lattice model is rather complicated as it captures the effects of magnetic monopoles. We note that \Rf{DW190608270} found that Eq.~\eq{CSlatt1} is invariant under the gauge transformation
\begin{align} 
\label{aZgauge} 
a^\RZ_I \to a^\RZ_I + m_I^\Z 
\end{align}
for any $\Z$-valued 1-cochain $m_I^\Z$ even when $\cB^3$ has boundary.

The path integral of the 3+1D bosonic model (for spacetime $\cM^4$ with or
without boundary) is then obtained from Eq.~\eq{CSlatt1} by taking a derivative and
setting ${g_3\to\infty}$. Using the properties of the (higher) cup product, the first line of Eq.~\eq{CSlatt1} vanishes since it is already the $\dd$ of something, the second line of Eq.~\eq{CSlatt1} becomes 
\begin{equation}
\begin{aligned}
& \ee^{\ii 2\pi  \int_{\cM^4} k_{IJ}\dd \big( a^\RZ_{I} (\dd a^\RZ_{J} -\toZ{\dd a^\RZ_J})-\toZ{\dd a^\RZ_I}a^\RZ_J \big)}
\\
&\hspace{15pt}=
\ee^{\ii 2\pi k_{IJ} \int_{\cM^4} (\dd a^\RZ_{I} -\toZ{\dd a^\RZ_I}) (\dd a^\RZ_{J} -\toZ{\dd a^\RZ_J}) }\times
\\
&\hspace{50pt}\ee^{\ii 2\pi k_{IJ} \int_{\cM^4} 
a^\RZ_{I} \dd \toZ{\dd a^\RZ_J}
- \dd \toZ{\dd a^\RZ_I}  a^\RZ_{J} 
},
\end{aligned}
\end{equation}
and the third line of Eq.~\eq{CSlatt1} becomes
\begin{equation}
\begin{aligned}
&\ee^{-\ii 2\pi k_{IJ} \int_{\cM^4} \dd \big( a^{\RZ}_J\hcup{1}\dd \toZ{\dd a^{\RZ}_I} \big) }\\
&\hspace{15pt}=\ee^{-\ii 2\pi  k_{IJ} \int_{\cM^4} \dd a^{\RZ}_J\hcup{1}\dd \toZ{\dd a^{\RZ}_I}}\times
\\
&\hspace{45pt}\ee^{\ii 2\pi k_{IJ} \int_{\cM^4} a^{\RZ}_J\dd \toZ{\dd a^{\RZ}_I} + \dd \toZ{\dd a^{\RZ}_I} a^{\RZ}_J } .
\end{aligned}
\end{equation}
Putting this all together and including the lattice Maxwell term ${ \ee^{- \sum_I \int_{\cM^4} \frac{|\dd
a^\RZ_I-\toZ{\dd a^\RZ_I}|^2}{g} } }$, we obtain a 3+1D bosonic model on spacetime lattice with a
$2\pi$-quantized topological term
\begin{equation}
\begin{aligned}
\label{K4Ddada}
&\hspace{-5pt}Z = \int D[a^\RZ_I] \
\ee^{-\ii 2\pi \hspace{-2pt}\sum\limits_{I\leq J} \hspace{-2pt}k_{IJ} \hspace{-4pt}\int\limits_{\cM^4} \hspace{-2pt}  \dd a^{\RZ}_J \hcup{1}\dd \toZ{\dd a^{\RZ}_I} }\\
&\hspace{7pt}\times
 \ee^{\ii 2\pi \hspace{-2pt}\sum\limits_{I\leq J} \hspace{-2pt}k_{IJ} \hspace{-4pt}\int\limits_{\cM^4} \hspace{-2pt} (\dd a^\RZ_I-\toZ{\dd a^\RZ_I})(\dd a^\RZ_J-\toZ{\dd a^\RZ_J})}\\
&\hspace{7pt}\times
\ee^{\ii 2\pi \hspace{-2pt}\sum\limits_{IJ}\hspace{-1pt}K_{IJ} \hspace{-4pt}\int\limits_{\cM^4} \hspace{-2pt}    a^{\RZ}_I  \dd \toZ{\dd a^{\RZ}_J} } 
\ee^{-  \hspace{-2pt}\sum\limits_{I}  \hspace{-2pt}\int\limits_{\cM^4} \hspace{-2pt} \frac{|\dd a^\RZ_I-\toZ{\dd a^\RZ_I}|^2}{g} }
\hspace{-2pt},
\end{aligned}
\end{equation}
where $K_{IJ}$ is given by
\begin{align}\label{kMatrix}
 K_{II}=2k_{II}, \ \ \ \ \ \
K_{IJ}=K_{JI}=k_{IJ}, \ \  I < J.
\end{align}

Because the lattice Chern-Simons path integral was invariant under the gauge
transformation Eq.~\eq{aZgauge} even when $\cM^4$ has boundary, by definition the path integral Eq.~\eqref{K4Ddada} is also invariant. Thus, requirement (1) from above is satisfied. Furthermore, since we defined the action as the derivative of something, requirement (2) is also automatically satisfied. Lastly, lets check that Eq.~\eqref{K4Ddada} satisfies requirement (3). In the ${g\sim 0}$ limit, the Maxwell term enforces fluctuations ${\dd a^\RZ_I-\toZ{\dd a^\RZ_I} \sim \eps}$ to be small. Therefore, using that
\begin{equation}
\dd \eps \sim 
\dd (\dd a^\RZ_I-\toZ{\dd a^\RZ_I})
=\dd \toZ{\dd a^\RZ_I},
\end{equation}
since $\dd \toZ{\dd a^\RZ_I}\se{1}0$ and $\eps$ is small, this implies that
\begin{equation}
 \dd \toZ{\dd a^\RZ_I} = 0,
\end{equation}
and hence there are no monopoles. When $
a^\RZ_J$ describes a monopole, it cannot be smooth and thus ${\dd\toZ{\dd
a^\RZ_J}\neq 0}$.  In fact, $\toZ{\dd a^\RZ_J}$ is the Poincar\'e dual of the
Dirac monopoles' worldsheets (\ie the trajectory of the Dirac strings of the monopole in
spacetime).  Thus $\dd \toZ{\dd a^\RZ_J}$ is the Poincar\'e dual of the
boundary of the Dirac worldsheet, which is the worldline of the $U(1)$
monopoles.

Therefore, the $g\sim 0$ limit corresponds to the smooth field limit. In this limit, the action amplitude for the topological term in Eq.~\eqref{K4Ddada} becomes
\begin{equation}\label{gTOzerotopactionamp}
\ee^{\ii 2\pi \hspace{-2pt}\sum\limits_{I\leq J} \hspace{-2pt}k_{IJ} \hspace{-4pt}\int\limits_{\cM^4} \hspace{-2pt} (\dd a^\RZ_I-\toZ{\dd a^\RZ_I})^2}.
\end{equation}
Relating the $1$-form field $a^I$ to the 1-cochain $(a^{\RZ}_I)_{ij}$ by
\begin{equation}
\int_i^j a^I = (a^{\RZ}_I)_{ij}
\end{equation}
and the 2-form curvature field ${f^I = \dd a^I}$ by
\begin{equation}
\int_{A(ijk)}\hspace{-8pt}f^I =  (\dd a^\RZ_I -\toZ{\dd a^\RZ_I})_{ijk},
\end{equation}
the action amplitude~\eqref{gTOzerotopactionamp} becomes
\begin{equation}\label{smallgTopTerm}
\ee^{\ii 2\pi \hspace{-2pt}\sum\limits_{I\leq J} \hspace{-2pt}k_{IJ} \hspace{-4pt}\int\limits_{\cM^4} \hspace{-2pt} (\dd a^\RZ_I-\toZ{\dd a^\RZ_I})^2}\approx\ee^{\ii 2\pi \hspace{-2pt}\sum\limits_{I\leq J}\hspace{-2pt}  k_{IJ} \hspace{-4pt} \int\limits_{M^4}  \hspace{-2pt} f^I\wdg f^J}.
\end{equation}
Therefore, in the smooth field limit (the low energy limit) the $2\pi$-quantized term on the lattice is captured by the continuum field theory and requirement (3) is satisfied.

In the absence of monopoles, Eq.~\eqref{smallgTopTerm} correctly becomes unity on a closed spacetime. For large $g$, however, due to the presence of monopoles the lattice term ${\ee^{\ii 2\pi \hspace{-2pt}\sum\limits_{I\leq J} \hspace{-2pt}k_{IJ} \hspace{-4pt}\int\limits_{\cM^4} \hspace{-2pt} (\dd
a^\RZ_I-\toZ{\dd a^\RZ_I})(\dd a^\RZ_J-\toZ{\dd a^\RZ_J})} }$ is no longer unity when $\cM^4$ is closed and thus is neither $2\pi$-quantized nor topological. 
Therefore, for large $g$ the low-energy limit of the lattice model may not be described by the continuum topological term~\eqref{contTopTerm1} since the lattice topological term must be described using all terms in Eq.~\eqref{K4Ddada}. It's more likely that the low-energy physics of the lattice model for large $g$, where the highly nontrivial terms in the first and third line of Eq.~\eqref{K4Ddada} are included, is better captured by the continuum topological term Eq.~\eqref{contTopTerm2}.

\subsection{1-symmetries in 3+1D $U^\ka(1)$ bosonic model}\label{sec:syms}

Now that we've introduced the $U^\ka(1)$ bosonic model, we now focus our attention on studying its symmetries and phase diagram. Firstly, let's review the case when $\cM^4$ has no boundary and the topological term vanishes and Eq.~\eq{K4Ddada} becomes Maxwell's theory
\begin{align}
\label{K4DdadaC}
Z(\cM^4) &= 
\int D[a^\RZ_I] \
\ee^{ -\sum_I \int_{\cM^4}\frac{|\dd a^\RZ_I-\toZ{\dd a^\RZ_I}|^2}{g} }.
\end{align}
When ${g\sim 0}$, the lattice curvature $\dd a^\RZ_I$ fluctuate weakly and so the above model
is in a deconfined phase of a compact $U^\ka(1)$ gauge theory and has a gapless photon excitation.
On the other hand, when ${g \to \infty}$ the model is in a gapped confined phase. Using that ${\int_{-\frac12}^{\frac12} \dd
(a^\RZ_I)_{ij} =1}$, the partition
function is
\begin{align}
Z(\cM^4) = \int D[a^\RZ_I]  =1,
\end{align}
for any closed spacetime $\cM^4$. According a conjecture in \Rf{KW1458}, this
implies that the gapped confined phase has a trivial topological order.

\begin{figure}[t]
\begin{center}
\includegraphics[width=.48\textwidth]{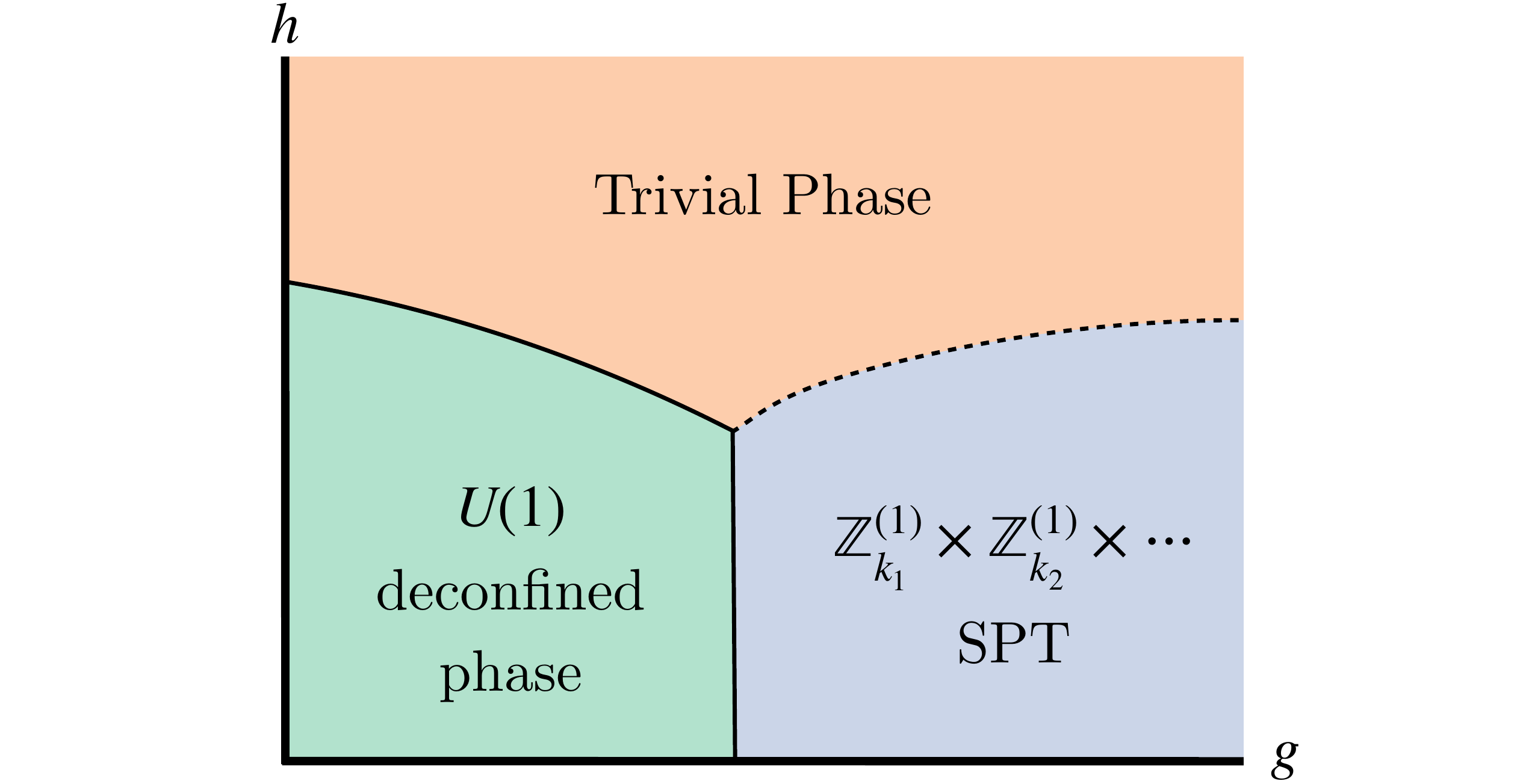}
\end{center}
\caption{The schematic phase diagram of the model described by Eq.~\eqref{K4Ddada} with
the additional term contributing to the action amplitude ${\ee^{h\sum_{ij,I}
a_{I,ij}^\RZ }}$.  When ${h=0}$, there is no fluctuations of $U(1)$ gauge
charge, and the model has an exact ${\Z_{k_1}^{(1)}\times \Z_{k_2}^{(1)}\times
\cdots}$ 1-symmetry.  For ${h\neq 0}$, this becomes an exact emergent
${\Z_{k_1}^{(1)}\times \Z_{k_2}^{(1)}\times \cdots}$ 1-symmetry existing below
the energy gaps of $U(1)$ gauge charges, which exists in the green and purple shaded regions. Due to the $2\pi$-quantized topological term, in the confined phase (shown in purple) there is an SPT order
protected by the exact emergent ${\Z_{k_1}^{(1)}\times \Z_{k_2}^{(1)}\times \cdots}$
1-symmetry whose SPT invariant is given by Eq.~\eqref{SPTinvkIJ}.} \label{U1ph} 
\end{figure}

In what follows, we now consider $\cM^4$ with a boundary so the $2\pi$-topological term contributes to the path integral. We'll show that the gapped confined phase now has a 1-SPT order
characterized by $k_{IJ}$ (see Fig. \ref{U1ph}). This is similar in spirit to section~\ref{Z2SPT} where in order to get $\Z_2^{(1)}$ SPT order we had to include the twist term Eq.~\eqref{ZbZ2t}.

Regardless the value of $g$ and even on $\cM^4$ with boundary, the path integral Eq.~\eqref{K4Ddada} is invariant under the transformation
\begin{align}
\label{1symm}
a^{\RZ}_{I} \to a^{\RZ}_{I} + \bt^{\QZ}_{I},\ \ \ \sum_I \bt^{\QZ}_{I}K_{IJ} \in \Z,\ \ \ \dd \bt^{\QZ}_{I}\se{1}0.
\end{align}
$\bt^{\QZ}_{I}$ are $\QZ$-valued 1-cocycles to ensure that the quantities ${\dd a^\RZ_I-\toZ{\dd
a^\RZ_I}}$ and ${\dd \toZ{\dd a^\RZ_I}}$ are invariant under the transformation
\eqref{1symm}. If this were the only requirement, Eq.~\eqref{1symm} would correspond to the $\kappa$ different $U(1)^{(1)}$ 1-symmetries. However, the additional constraint that $\sum_I
\bt^{\QZ}_{I}K_{IJ}$ are $\Z$-valued cochains is required when there are magnetic monopoles to ensure the term $\ee^{\ii 2\pi
\int_{\cM^4} \sum_{IJ}   a^{\RZ}_I K_{IJ} \dd \toZ{\dd a^{\RZ}_J} } $ is
invariant. Indeed, under the transformation Eq.~\eq{1symm},
this term changes by a phase factor 
\begin{equation*}
\ee^{\ii 2\pi\int_{\cM^4}\sum_{IJ}\bt^{\QZ}_I K_{IJ} \dd \toZ{\dd a^{\RZ}_J}},
\end{equation*}
which is $1$ provided $\bt^{\QZ}_{I}$ satisfy
${\sum_I \bt^{\QZ}_{I}K_{IJ} \in \Z}$. 

As an integer matrix, $K$ has the following Smith normal form
\begin{align}
\label{smith}
 K = U 
\begin{pmatrix}
 k_1 & & & \\
  &  k_2 & & \\
  &   &   k_3 & \\
  &   &    & \ddots \\
\end{pmatrix}
V ,
\end{align}
where $k_I$ are integers and $U,V$ are invertible integer matrices. 
Now the 1-symmetry can be written as
\begin{align}
\label{1symm0}
& a^{\RZ}_{I} \to 
a^{\RZ}_{I} + \bt^{\QZ}_{I} 
=
a^{\RZ}_{I} + \sum_J \t \bt^{\QZ}_{J} (U^{-1})_{JI},
\nonumber\\
& \sum_I \bt^{\QZ}_{I} U_{IJ}k_J 
= \t \bt^{\QZ}_{J}k_J
\in \Z,\ \ \ \dd \t \bt^{\QZ}_{I}\se{1}0.
\end{align}
We see that the 1-symmetry is a $\Z_{k_1}^{(1)}\times \Z_{k_2}^{(1)} \times
\cdots$ 1-symmetry generated by the quantized $\t \bt^{\QZ}_{J}$.  When
$k_I=0$, $\t \bt^{\QZ}_{I}$ is not quantized and generates $U(1)^{(1)}$
1-symmetry. The above result remains valid if we regard
$\Z_{0}^{(1)}$ as the $U(1)^{(1)}$ 1-symmetry.  We note that, since the
${\Z_{k_1}^{(1)}\times \Z_{k_2}^{(1)}\times \cdots}$ 1-symmetry is valid on the
spacetime lattice with or without boundary, the 1-symmetry is anomaly-free. 

In addition to giving rise to a finite 1-symmetry, the term $\ee^{\ii 2\pi \int_{\cM^4} \sum_{IJ}   a^{\RZ}_I
K_{IJ} \dd \toZ{\dd a^{\RZ}_J} } $ also causes the $U(1)$ monopoles to be bounded with the $U(1)$ charges.  In particular, the
unit monopole of the $J^\text{th}$ $U(1)$ field carries the $I^\text{th}$
$U(1)$ charge $K_{IJ}$. This is precisely the lattice version of the generalized Witten effect discussed in the continuum theory (see Eq.~\eqref{GWlaw}).

For large $g$, these monopole-charge bound states condense which gives rise to
a gapped oblique confined phase with $\Z_{k_1}^{(1)}\times \Z_{k_2}^{(1)}\times
\cdots$ 1-symmetry. We note that the 2+1D lattice $U^\ka(1)$ Chern-Simons
theory \eqref{CSlatt1} also has the $\Z_{k_1}^{(1)}\times \Z_{k_2}^{(1)}\times
\cdots$ 1-symmetry, which can actually be anomalous~\cite{DW190608270}. Since
the 2+1D lattice $U^\ka(1)$ Chern-Simons theory is the boundary of the
$U^\ka(1)$ model in the gapped confined phase, from the point of view of
anomaly inflow~\cite{KT13094721,W1313} the gapped confined phase may have a
non-trivial ${\Z_{k_1}^{(1)}\times \Z_{k_2}^{(1)}\times \cdots}$ 1-SPT order.
Indeed, in the next section we'll show that this confined phase is
characterized by the $K$-matrix and has a 1-SPT order protected by the
${\Z_{k_1}^{(1)}\times \Z_{k_2}^{(1)}\times \cdots}$ 1-symmetry. Indeed, the
1-SPT invariant found in the next section is given by Eq.~\eq{1SPTinvk}.

Before concluding this subsection, we remark that the fact that the
${\Z_{k_1}^{(1)}\times \Z_{k_2}^{(1)}\times \cdots}$ 1-symmetry is exact in the
bosonic model is a special feature of the theory. A more generic lattice theory
would also include the action amplitude ${\ee^{h\sum_{ij,I} a_{I,ij}^\RZ }}$ in
the path integral, which explicitly breaks the ${\Z_{k_1}^{(1)}\times
\Z_{k_2}^{(1)}\times \cdots}$ 1-symmetry. However, like in the $\Z_2$ gauge
theory case discussed in section~\ref{emergentZ21SPT}, for energies below the
$U(1)$ gauge charge gaps, there is a region of $h\neq 0$ where the
${\Z_{k_1}^{(1)}\times \Z_{k_2}^{(1)}\times \cdots}$ 1-symmetry is an exact emergent
symmetry. In this region, the corresponding 1-SPT order would also affect the
low-energy physics and be protected by the exact emergent symmetry (see
Fig.~\ref{U1ph}).

\subsection{Gauging the ${\Z_{k_1}^{(1)}\times \Z_{k_2}^{(1)}\times ... }$
1-symmetry}\label{gaugeSym}

The fact that the boundary Chern-Simons theory has an anomalous
${\Z_{k_1}^{(1)}\times \Z_{k_2}^{(1)}\times \cdots}$
1-symmetry~\cite{DW190608270} means that the bulk theory has 1-SPT order  in
the large-$g$ confined phase protected by ${\Z_{k_1}^{(1)}\times
\Z_{k_2}^{(1)}\times \cdots}$. In this section, we will characterize the 1-SPT
theory in the bulk 3+1D theory by finding the SPT invariant obtained by gauging
the 1-symmetry. This subsection contains mostly detailed calculations in order
to derive the 1-SPT invariant given by Eq.~\eqref{1SPTinvk}.

Before gauging the symmetry, it's convenient to first slowly turn on addition
terms in the action which will not affect the 1-SPT order. In particularly, to
the Euclidean lattice action we add
\begin{align}\label{Uterm}
S \supset U \sum_{ij,J} \cos\left(2\pi
\sum_I (a^{\RZ}_{I})_{ij} K_{IJ} \right).
\end{align}
Note that in the ${U\to\infty}$ limit, this term makes $a^\RZ_I$ satisfy the quantization condition
\begin{align}
 \sum_I a^{\RZ}_{I}K_{IJ} \se{1} 0.
\label{aq}
\end{align}
Crucially, this preserves the ${\Z_{k_1}^{(1)}\times \Z_{k_2}^{(1)}\times \cdots}$ 1-symmetry whose transformation is Eq.~\eqref{1symm}. Furthermore, when ${g\to\infty}$ the path integral for any closed spacetime changes smoothly as $U$ changes from $0$ to $\infty$. This is because in this limit the only term in the action is Eq.~\eqref{Uterm} which is independently defined on each 1-simplex of the spacetime triangulation (i.e., non-interacting). Thus, the $U=0$ state and the ${U\to\infty}$ state belong to the same phase and so the ${(g,U)=(\infty,\infty)}$ phase has the 1-SPT order as the ${(g,U)=(\infty,0)}$ phase. By considering the $U\to\infty$ state, the quantization condition turns the $U(1)$ cochain fields $a^{\RZ}_I$ into discrete cochain fields ${a^\QZ_I}$\footnote{${a^\RZ_I}$ is renamed as ${a^\QZ_I}$, since the quantized ${a^\QZ_I}$'s, ${\sum_I a^\QZ_{I} K_{IJ} \se{1}0}$, have values in $\QZ$.} satisfying Eq. \eq{aq}, which allows us to use results and techniques for discrete fields from section~\ref{Z2SPT} to study the 1-SPT order in the $U(1)$ model.

We now consider the ${U\to\infty}$ state, which in the strongly-interacting limit ${g\to\infty}$ the path integral Eq.~\eq{K4Ddada} becomes
\begin{equation}\label{K4DdadaD}
\begin{aligned}
Z &= 
\hskip -1.5em 
\sum_{\sum_I a^\QZ_{I} K_{IJ} \se{1}0}
\hskip -1.5em 
\ee^{-\ii 2\pi \hspace{-2pt}\sum\limits_{I\leq J} \hspace{-2pt}k_{IJ} \hspace{-4pt}\int\limits_{\cM^4} \hspace{-2pt}  
\dd a^{\QZ}_J \hcup{1}\dd \toZ{\dd a^{\QZ}_I} }\times\\
&
\hspace{5pt} \ee^{\ii 2\pi \hspace{-2pt}\sum\limits_{I\leq J} \hspace{-2pt}k_{IJ} \hspace{-4pt}\int\limits_{\cM^4} \hspace{-2pt} (\dd a^\QZ_I-\toZ{\dd a^\QZ_I})(\dd a^\QZ_J-\toZ{\dd a^\QZ_J})},
\end{aligned}
\end{equation}
where we have use that for quantized $a^\QZ_I$ satisfying Eq.~\eqref{aq},
\begin{align}
\ee^{\ii 2\pi \int_{\cM^4} \sum_{IJ} a^{\RZ}_I K_{IJ} \dd \toZ{\dd a^{\RZ}_J} }
=1.
\end{align} 
As mentioned, just like the original path integral Eq.~\eqref{K4Ddada}, this path integral Eq.~\eqref{K4DdadaD} also has the anomaly-free
1-symmetry
\begin{align}
\label{1symmD}
a^{\QZ}_{I} \to a^{\QZ}_{I} + \bt^{\QZ}_{I},\ \ \ \sum_I \bt^{\QZ}_{I}K_{IJ} \in \Z,\ \ \ \dd \bt^{\QZ}_{I}\se{1}0.
\end{align}
Lastly, using that ${\toZ{\dd a^{\RZ}_J}\hcup{1}\dd\toZ{\dd a^{\RZ}_I}}\in\Z$ and ${\dd(\dd a^{\RZ}_I) = 0}$, we insert unity of the form
\begin{equation}\label{unityInsert}
\begin{aligned}
1 &= \ee^{\ii2\pi \hspace{-2pt}\sum\limits_{I\leq J} \hspace{-2pt}k_{IJ} \hspace{-4pt}\int\limits_{\cM^4} \hspace{-2pt}\toZ{\dd a^{\RZ}_J}\hcup{1}\dd\toZ{\dd a^{\RZ}_I}}\times\\
&\hspace{20pt}\ee^{\ii2\pi \hspace{-2pt}\sum\limits_{I\leq J} \hspace{-2pt}k_{IJ} \hspace{-4pt}\int\limits_{\cM^4} \hspace{-2pt}(\dd a^{\RZ}_J-\toZ{\dd a^{\RZ}_J})\hcup{1}\dd(\dd a^{\RZ}_I)}
\end{aligned}
\end{equation}
into Eq.~\eqref{K4DdadaD} such that the path integral becomes
\begin{equation}\label{K4DdadaD1}
\begin{aligned}
Z &= 
\hskip -2em 
\sum_{\sum_I a^\QZ_{I} K_{IJ} \se{1}0}
\hskip -2em 
\ee^{\ii 2\pi \hspace{-2pt}\sum\limits_{I\leq J} \hspace{-2pt}k_{IJ} \hspace{-4pt}\int\limits_{\cM^4} \hspace{-2pt} 
( \dd a^{\QZ}_J -\toZ{\dd a^{\QZ}_J}) 
\hcup{1}\dd (\dd a^{\QZ}_I-\toZ{\dd a^{\QZ}_I} )}\times\\
&
\hspace{15pt} \ee^{\ii 2\pi \hspace{-2pt}\sum\limits_{I\leq J} \hspace{-2pt}k_{IJ} \hspace{-4pt}\int\limits_{\cM^4} \hspace{-2pt} (\dd a^\QZ_I-\toZ{\dd a^\QZ_I})(\dd a^\QZ_J-\toZ{\dd a^\QZ_J})}.
\end{aligned}
\end{equation}

To determine the SPT order realized by the theory Eq.~\eqref{K4DdadaD}, we gauge the 1-symmetries by first replacing $\dd a^\QZ_I$ with ${\dd a^\QZ_I -  B^\QZ_I}$ which for convenience we'll denote as
\begin{align}
b_I^\QZ & \equiv \dd a^\QZ_I -  B^\QZ_I, 
\end{align}
where $B^\QZ_I$ is a background symmetry twist field satisfying
\begin{equation}
\dd  B^\QZ_I  \se{1} 0, \quad\quad
\sum_{I}   B^\QZ_I K_{IJ} \se{1} 0.
\end{equation}
We use ``$\se{1}$'' instead of ``$=$'' here since shifting $B_I^\QZ$ by a $\Z$-valued 2-cochain corresponds to performing a gauge transformation. After this, the path-integral Eq.~\eq{K4DdadaD1} of course becomes
\begin{equation}\label{Zab2}
\begin{aligned}
Z &= 
\hskip -2em 
\sum_{\sum_I a^\QZ_{I} K_{IJ} \se{1}0}
\hskip -2em 
\ee^{\ii 2\pi \hspace{-2pt}\sum\limits_{I\leq J} \hspace{-2pt}k_{IJ} \hspace{-4pt}\int\limits_{\cM^4} \hspace{-2pt} 
( b^{\QZ}_J -\toZ{b^{\QZ}_J}) 
\hcup{1}\dd (b^{\QZ}_I-\toZ{b^{\QZ}_I} )}\times\\
&
\hspace{15pt} \ee^{\ii 2\pi \hspace{-2pt}\sum\limits_{I\leq J} \hspace{-2pt}k_{IJ} \hspace{-4pt}\int\limits_{\cM^4} \hspace{-2pt} (b^\QZ_I-\toZ{b^\QZ_I})(b^\QZ_J-\toZ{b^\QZ_J})}.
\end{aligned}
\end{equation}
Note that for $\cM^4$ with or without boundary, Eq.~\eqref{Zab2} is importantly invariant under the gauge transformations
\begin{equation}\label{abZgauge}
\begin{aligned}
 a_I^\QZ &\to a_I^\QZ + m^\Z_I,\\
 b_I^\QZ &\to b_I^\QZ + n^\Z_I,
\end{aligned}
\end{equation}
where ${m^\Z_I\se{1}0}$ and ${n^\Z_I \se{1}0}$.
If we had not inserted Eq.~\eqref{unityInsert} into Eq.~\eqref{K4DdadaD}, the gauged theory would have not been gauge invariant.

The path integral Eq.~\eqref{Zab2} is a bit cumbersome in its current form and it's hard to see how $\cM^4$ being opened or closed changes the action amplitude. Thus, let's massage the action amplitude of Eq.~\eqref{Zab2} a bit to get it in a more enlightening form. First, we consider the first line of Eq.~\eqref{Zab2}. Using that ${\dd b_I^{\QZ} \se{1} 0}$ and rewriting
\begin{equation}
\begin{aligned}
&
\ee^{\ii 2\pi \hspace{-2pt}\sum\limits_{I\leq J} \hspace{-2pt}k_{IJ} \hspace{-4pt}\int\limits_{\cM^4} \hspace{-2pt}
(b^{\QZ}_J -\toZ{b^{\QZ}_J})
\hcup{1}\dd(b^{\QZ}_I - \toZ{b^{\QZ}_I} )}
\\
&
\hspace{20pt}=\ee^{\ii 2\pi \hspace{-2pt}\sum\limits_{I\leq J} \hspace{-2pt}k_{IJ} \hspace{-4pt}\int\limits_{\cM^4} \hspace{-2pt}
b^{\QZ}_J \hcup{1}\dd(b^{\QZ}_I - \toZ{b^{\QZ}_I} )},
\end{aligned}
\end{equation}
we then can use Eq.~\eqref{cupkrel} and once again ${\dd b_I^{\QZ} \se{1} 0}$ to write this as
\begin{equation}
\begin{aligned}
&
\ee^{\ii 2\pi \sum\limits_{I\leq J} k_{IJ}  \hspace{-4pt}\int\limits_{\cM^4} \hspace{-2pt}
b^{\QZ}_J \hcup{1}\dd(b^{\QZ}_I - \toZ{b^{\QZ}_I} )}\\
&\hspace{5pt}= \ee^{\ii 2\pi \sum\limits_{I\leq J} k_{IJ}  \hspace{-4pt}\int\limits_{\cM^4} \hspace{-2pt}\dd  \big(b^{\QZ}_J \hcup{1} (b^{\QZ}_I -\toZ{b^{\QZ}_I})  \big) -\dd b^{\QZ}_J \hcup{1} b^{\QZ}_I }\times\\
&\hspace{25pt}
\ee^{\ii 2\pi \sum\limits_{I\leq J} k_{IJ}  \hspace{-4pt}\int\limits_{\cM^4} \hspace{-2pt}
\toZ{b^{\QZ}_I} b^{\QZ}_J  -b^{\QZ}_J \toZ{b^{\QZ}_I}
} .
\end{aligned}
\end{equation}
Next, we consider the second line of Eq.~\eqref{Zab2}. We can use the fact that since ${\toZ{b_I^\QZ}\toZ{b_J^\QZ}\in\Z}$, then
\begin{equation}
\begin{aligned}
&\ee^{\ii 2\pi \hspace{-2pt}\sum\limits_{I\leq J} \hspace{-2pt}k_{IJ} \hspace{-4pt}\int\limits_{\cM^4} \hspace{-2pt} (b_I^\QZ-\toZ{b_I^\QZ})(b_J^\QZ-\toZ{b_J^\QZ}) }\\
&\hspace{20pt}= \ee^{\ii 2\pi \hspace{-2pt}\sum\limits_{I\leq J} \hspace{-2pt}k_{IJ} \hspace{-4pt}\int\limits_{\cM^4} \hspace{-2pt} b_I^\QZ b_J^\QZ -b_I^\QZ\toZ{b_J^\QZ} -\toZ{b_I^\QZ}b_J^\QZ }.
\end{aligned}
\end{equation}
Using these simplifications, the gauged model \eqref{Zab2} can be rewritten as
\begin{align}
\label{Zab}
Z = 
\hskip -1.5em 
\sum_{\sum_I a^\QZ_{I} K_{IJ} \se{1}0}
\hskip -1.5em 
 & \ee^{\ii 2\pi \hspace{-2pt}\sum\limits_{I\leq J} \hspace{-2pt}k_{IJ} \hspace{-4pt}\int\limits_{\cM^4} \hspace{-2pt} b_J^\QZ b_I^\QZ
-\dd b^{\QZ}_J \hcup{1} b^{\QZ}_I } 
\\
& 
 \ee^{\ii 2\pi \hspace{-2pt}\sum\limits_{I\leq J} \hspace{-2pt}k_{IJ} \hspace{-4pt}\int\limits_{\cM^4} \hspace{-2pt} 
\dd\big(b^{\QZ}_J \hcup{1}\dd(b^{\QZ}_I - \toZ{b^{\QZ}_I}) \big) },
\nonumber 
\end{align}
where we have used that ${\sum_{IJ}   b^\QZ_I K_{IJ} \se{1} 0 }$,
\begin{align}
\ee^{\ii 2\pi \hspace{-2pt}\sum\limits_{I\leq J} \hspace{-2pt}k_{IJ} \hspace{-4pt}\int\limits_{\cM^4} \hspace{-2pt} 
-b_I^\QZ\toZ{b_J^\QZ} 
-b_J^\QZ\toZ{b_I^\QZ} 
}
=1.
\end{align}
Therefore, starting from the $U^{\kappa}(1)$ bosonic model and gauging the ${\Z_{k_1}^{(1)}\times \Z_{k_2}^{(1)}\times \cdots}$, Eq.~\eqref{Zab} gives the path integral of the gauged model from which we can find the 1-SPT invariant.

When the spacetime $\cM^4$ has no boundary, the total derivative term in Eq.~\eqref{Zab} vanishes and the path integral becomes
\begin{equation}\label{Zab0}
Z(B^\QZ_I)  = 
\hskip -1.7em 
\sum_{\sum_I a^\QZ_{I} K_{IJ} \se{1}0}
\hskip -1.7em 
 \ee^{\ii 2\pi \hspace{-2pt}\sum\limits_{I\leq J} \hspace{-2pt}k_{IJ} \hspace{-4pt}\int\limits_{\cM^4} \hspace{-2pt} b_J^\QZ b_I^\QZ
-\dd b^{\QZ}_J \hcup{1} b^{\QZ}_I } .
\end{equation}
We note that Eq.~\eqref{Zab0} is
invariant under the following gauge transformation:
\begin{align}
\label{bgauge}
 b^{\QZ}_I \to b^{\QZ}_I +\dd \omega^{\QZ}_I,
\ \ \ \
\sum_I \omega^\QZ_{I} K_{IJ} \se{1}0.
\end{align}
It is straight forward to check that this is indeed the case. When ${\prt \cM^4 =\emptyset}$, the gauge transformation Eq.~\eq{bgauge} changes the
term ${\ee^{\ii 2\pi \sum_{I\leq J}k_{IJ} \int_{\cM^4} \hspace{-2pt} b_J^\QZ b_I^\QZ} }$ by a
factor 
\begin{equation}\label{gTransPickUp1}
\begin{aligned}
&\ee^{\ii 2\pi \hspace{-2pt}\sum\limits_{I\leq J} \hspace{-2pt}k_{IJ} \hspace{-4pt}\int\limits_{\cM^4} \hspace{-2pt} \dd \omega_J^\QZ b_I^\QZ + b_J^\QZ \dd \omega_I^\QZ +\dd \omega_J^\QZ \dd \omega_I^\QZ} \\
&\hspace{20pt}= \ee^{\ii 2\pi \hspace{-2pt}\sum\limits_{I\leq J} \hspace{-2pt}k_{IJ} \hspace{-4pt}\int\limits_{\cM^4} \hspace{-2pt} \omega_J^\QZ \dd b_I^\QZ -\dd b_J^\QZ \omega_I^\QZ }.
\end{aligned}
\end{equation}
However, using that ${\ee^{\ii 2\pi \sum_{I, J} K_{IJ} \int_{\cM^4} \omega_J^\QZ \dd b_I^\QZ } =1}$ from Eq.~\eqref{bgauge}, and also Eq.~\eq{cupkrel}, we can rewrite Eq.~\eqref{gTransPickUp1} as
\begin{equation}\label{gTransPickUp1b}
\ee^{-\ii 2\pi \hspace{-2pt}\sum\limits_{I\leq J} \hspace{-2pt}k_{IJ} \hspace{-4pt}\int\limits_{\cM^4} \hspace{-2pt} 
\dd \omega_I^\QZ \hcup{1} \dd b_J^\QZ } .
\end{equation}
Furthermore, the gauge transformation \eqref{bgauge} changes the term
$ {\ee^{\ii 2\pi \sum_{I\leq J} k_{IJ} \int_{\cM^4} b^{\QZ}_I \hcup{1} \dd b^{\QZ}_J } }
$
by a factor
\begin{align}\label{gTransPickUp2}
\ee^{\ii 2\pi \hspace{-2pt}\sum\limits_{I\leq J} \hspace{-2pt}k_{IJ} \hspace{-4pt}\int\limits_{\cM^4} \hspace{-2pt} \dd \omega^{\QZ}_I \hcup{1} \dd b^{\QZ}_J } .
\end{align}
Eq.~\eqref{gTransPickUp1b} and~\eqref{gTransPickUp2} perfectly cancel each other and, therefore, the action amplitude in
Eq.~\eq{Zab0} is gauge invariant.

Because the action amplitude is invariant under Eq.~\eqref{bgauge}, it will not depend on the coboundaries $\dd a_I^\QZ$ and, therefore, we will be able to evaluate the path
integral Eq.~\eq{Zab} when $\prt \cM^4=\emptyset$. Plugging in $b_I^\QZ$ and integrating by parts using that  $\cM^4$ is closed, the path integral Eq.~\eqref{Zab0} becomes
\begin{equation*}
\begin{aligned}
Z  &= 
\hskip -1.7em 
\sum_{\sum_I a^\QZ_{I} K_{IJ} \se{1}0}
\hskip -1.7em 
 \ee^{\ii 2\pi \hspace{-2pt}\sum\limits_{I\leq J} \hspace{-2pt}k_{IJ} \hspace{-4pt}\int\limits_{\cM^4} \hspace{-2pt} -a_J^\QZ\dd B_I^\QZ + \dd B_J^\QZ a_I^\QZ}\times\\
&
\ee^{\ii 2\pi \hspace{-2pt}\sum\limits_{I\leq J} \hspace{-2pt}k_{IJ} \hspace{-4pt}\int\limits_{\cM^4} \hspace{-2pt} B_J^\QZ B_I^\QZ + \dd B_J^\QZ \hcup{1} \dd a_I^\QZ
- \dd B_J^\QZ \hcup{1} B_I^\QZ}.
\end{aligned}
\end{equation*}
Now, we can use Eq.~\eqref{cupkrel} to rewrite the terms ${a_J^\QZ\dd B_I^\QZ}$ and ${\dd B_J^\QZ a_I^\QZ}$ such that $Z$ becomes
\begin{equation*}
\begin{aligned}
Z  &= 
\hskip -1.7em 
\sum_{\sum_I a^\QZ_{I} K_{IJ} \se{1}0}
\hskip -1.7em 
\ee^{\ii 2\pi \hspace{-2pt}\sum\limits_{I\leq J} \hspace{-2pt}k_{IJ} \hspace{-4pt}\int\limits_{\cM^4} \hspace{-2pt} B_J^\QZ B_I^\QZ - \dd B_J^\QZ \hcup{1} B_I^\QZ}\times\\
&\hspace{30pt}
\ee^{-\ii 2\pi \hspace{-2pt}\sum\limits_{I,J} \hspace{-2pt}K_{IJ} \hspace{-4pt}\int\limits_{\cM^4} \hspace{-2pt} a_I^\QZ \dd B_J^\QZ }.
\end{aligned}
\end{equation*}
Because the path integral only sums over $a_I^\QZ$ satisfying the quantization condition ${\sum_I a^\QZ_{I} K_{IJ} \se{1}0}$ and that ${\dd B_J^\QZ \se{1} 0}$, the term in the second line of $Z$ becomes unity. Then, using Eq.~\eqref{cupkrel} to rewrite ${\dd B_J^\QZ \hcup{1} B_I^\QZ}$, the path integral becomes
\begin{equation*}
\begin{aligned}
Z &= 
\hskip -1.7em 
\sum_{\sum_I a^\QZ_{I} K_{IJ} \se{1}0}
\hskip -1.7em 
\ee^{\ii 2\pi \hspace{-2pt}\sum\limits_{I\leq J} \hspace{-2pt}k_{IJ} \hspace{-4pt}\int\limits_{\cM^4} \hspace{-2pt} B_J^\QZ B_I^\QZ + B_I^\QZ \hcup{1} \dd B_J^\QZ }\times\\
&\hspace{30pt}
\ee^{-\ii 2\pi \hspace{-2pt}\sum\limits_{I,J} \hspace{-2pt}K_{IJ} \hspace{-4pt}\int\limits_{\cM^4} \hspace{-2pt} \dd B_J^\QZ\hcup{2} \dd B_J^\QZ }.
\end{aligned}
\end{equation*}
Firstly, note that the action amplitude on the second line is unity since ${\dd B_J^\QZ\hcup{2} \dd B_J^\QZ\in\Z}$. Additionally, the action amplitude no longer contains the cochains $a_I^\QZ$ which the path integral is summing over. Thus, performing the sum we obtain 
\begin{equation}
Z \hspace{-1pt}= \hspace{-1pt}|\det(K)|^{N_e}~
 \ee^{\ii 2\pi \hspace{-2pt}\sum\limits_{I\leq J} \hspace{-2pt}k_{IJ} \hspace{-4pt}\int\limits_{\cM^4} \hspace{-2pt}  B_J^\QZ  B_I^\QZ
+  B^{\QZ}_I \hcup{1} \dd  B^{\QZ}_J } \hspace{-6pt}.
\end{equation}
where $N_e$ is the number of edges in the triangulated spacetime $\cM^4$.

\subsection{The 1-SPT invariant}\label{section:SPTinv}

The SPT order is characterized by the
volume-independent partition function
\begin{equation}
Z^{\text{top}}(\cM^4,B_I^{\QZ}) 
=  \dfrac{Z (\cM^4,B_I^{\QZ})}{Z (\cM^4,0)}.
\end{equation}
From this, we find that 1-SPT invariant for the 1-SPT state is
\begin{equation}\label{SPTinvkIJ}
Z^\text{top}(\cM^4, B^\QZ_I) = 
 \ee^{\ii 2\pi \hspace{-2pt}\sum\limits_{I\leq J} \hspace{-2pt}k_{IJ} \hspace{-4pt}\int\limits_{\cM^4} \hspace{-2pt}  B_J^\QZ  B_I^\QZ
+  B^{\QZ}_I \hcup{1} \dd  B^{\QZ}_J } ,
\end{equation}
where as a reminder
\begin{equation}
\dd  B^{\QZ}_I \se{1} 0, \ \ \ \ \
\sum_I B^\QZ_{I} K_{IJ} \se{1}0.
\end{equation}
Such a non-trivial 1-SPT invariant Eq.~\eqref{SPTinvkIJ} suggests that the 1-SPT order can be non-trivial. We note that, as confirmed in appendix section~\ref{BgaugeInv}, this 1-SPT invariant is correctly gauge invariant.

However, before going on consider some examples of non-trivial 1-SPT invariants
(see section~\ref{examplesSec}), we want to show that any two matrices $K$ and
$\t K$ related by ${\t K =U^\top K U}$ with ${U\in GL(\ka,\Z)}$ actually
describes the same 1-SPT invariant.  We will first try to express the 1-SPT
invariant Eq.~\eqref{SPTinvkIJ} in terms of only the $K$-matrix instead of
$k_{IJ}$. In doing so, we'll also find a nice form for the 1-SPT invariant
which we can use when considering examples in the next section. 

Consider the term ${B_J^\QZ  B_I^\QZ}$ in the SPT invariant Eq.~\eqref{SPTinvkIJ}. We can first rewrite it as
\begin{equation*}
\begin{aligned}
\sum\limits_{I\leq J} \hspace{-2pt}k_{IJ} \hspace{-4pt}\int\limits_{\cM^4} \hspace{-2pt}  B_J^\QZ  B_I^\QZ &=\frac12 \hspace{-2pt}\sum\limits_{I\leq J} \hspace{-2pt}k_{IJ} \hspace{-4pt}\int\limits_{\cM^4} \hspace{-2pt}  B_J^\QZ  B_I^\QZ - B_I^\QZ  B_J^\QZ\\
&\hspace{15pt}+\frac12 \hspace{-2pt}\sum\limits_{I,J} \hspace{-2pt}K_{IJ} \hspace{-4pt}\int\limits_{\cM^4} \hspace{-2pt}  B_I^\QZ  B_J^\QZ.\\
\end{aligned}
\end{equation*}
Then using Eq.~\eq{cupkrel} and the fact that $\cM^4$ is closed, this can become
\begin{equation*}
\begin{aligned}
&\frac12 \hspace{-2pt}\sum\limits_{I\leq J} \hspace{-2pt}k_{IJ} \hspace{-4pt}\int\limits_{\cM^4} \hspace{-2pt}  \dd B_J^\QZ \hcup{1} B_I^\QZ 
+ B_J^\QZ\hcup{1} \dd B_I^\QZ
\\
&\ \ \ \ 
+\frac12 \sum_{I,J} K_{IJ} \int_{\cM^4}  B_I^\QZ  B_J^\QZ, \\
&=\frac12 \hspace{-2pt}\sum\limits_{I\leq J} \hspace{-2pt}k_{IJ} \hspace{-4pt}\int\limits_{\cM^4} \hspace{-2pt}  \dd B_J^\QZ \hcup{1} B_I^\QZ 
- B_I^\QZ\hcup{1} \dd B_J^\QZ\\
&\ \ \ \ 
+\frac12 \sum_{I,J} K_{IJ} \int_{\cM^4}  B_I^\QZ  B_J^\QZ + B^{\QZ}_I \hcup{1} \dd  B^{\QZ}_J.
\end{aligned}
\end{equation*}
Plugging this expression for ${B_J^\QZ  B_I^\QZ}$ into the action in Eq.~\eqref{SPTinvkIJ}, we find
\begin{equation*}
\begin{aligned}
 &\sum\limits_{I\leq J} \hspace{-2pt}k_{IJ} \hspace{-4pt}\int\limits_{\cM^4} \hspace{-2pt}  B_J^\QZ  B_I^\QZ + B^{\QZ}_I \hcup{1} \dd  B^{\QZ}_J
\\
&=\frac12 \sum_{I,J} K_{IJ} \int_{\cM^4}  B_I^\QZ  B_J^\QZ + B^{\QZ}_I \hcup{1} \dd  B^{\QZ}_J\\
&\ \ \ \ 
+\frac12 \hspace{-2pt}\sum\limits_{I\leq J} \hspace{-2pt}k_{IJ} \hspace{-4pt}\int\limits_{\cM^4} \hspace{-2pt}  \dd B_J^\QZ \hcup{1} B_I^\QZ 
+ B_I^\QZ\hcup{1} \dd B_J^\QZ 
.
\end{aligned}
\end{equation*}
We can go further by again using Eq.~\eq{cupkrel} to rewrite second line of the right hand side and get
\begin{equation}
\begin{aligned}
&\sum\limits_{I\leq J} \hspace{-2pt}k_{IJ} \hspace{-4pt}\int\limits_{\cM^4} \hspace{-2pt}  B_J^\QZ  B_I^\QZ + B^{\QZ}_I \hcup{1} \dd  B^{\QZ}_J\\
&= \frac12 \sum_{I,J} K_{IJ} \int_{\cM^4}  B_I^\QZ  B_J^\QZ + B^{\QZ}_I \hcup{1} \dd  B^{\QZ}_J\\
&\ \ \ \ 
-\frac12 \hspace{-2pt}\sum\limits_{I\leq J} \hspace{-2pt}k_{IJ} \hspace{-4pt}\int\limits_{\cM^4} \hspace{-2pt}  \dd B_J^\QZ \hcup{2} \dd B_I^\QZ 
 .
\end{aligned}
\end{equation}
Therefore, the 1-SPT invariant that characterizes the 1-SPT order has the form
\begin{equation}\label{1SPTinv}
\begin{aligned}
\hspace{-5pt}Z^\text{top}(\cM^4, B^\QZ_I\hspace{-1pt}) &\hspace{-2pt}=\hspace{-2pt} 
\ee^{\ii \pi \hspace{-2pt}\sum\limits_{I, J} \hspace{-2pt}K_{IJ} \hspace{-4pt}\int\limits_{\cM^4} \hspace{-2pt} B_I^\QZ  B_J^\QZ + B^{\QZ}_I \hcup{1} \dd  B^{\QZ}_J}  \\
&\ \ \ \ \times\ee^{\ii \pi  
\hspace{-2pt}\sum\limits_{I\leq J} \hspace{-2pt}k_{IJ} \hspace{-4pt}\int\limits_{\cM^4} \hspace{-2pt} \dd B_I^\QZ \hcup{2} \dd B_J^\QZ
} .
\end{aligned}
\end{equation}
We can recast the relationship between $k_{IJ}$ and $K_{IJ}$, given by Eq.~\eqref{kMatrix}, by treating $k_{IJ}$ as the elements of the upper triangular integer matrix $k$ that satisfies
\begin{align}
 K=k+k^\top .
\end{align}
Then, using that ${\dd B_J^\QZ \hcup{2} \dd B_I^\QZ \se{\dd} -
\dd B_I^\QZ \hcup{2} \dd B_J^\QZ}$ from Eq. \eq{cupkrel}, we can replace ${\sum_{I\leq J}
k_{IJ}}$ by ${\sum_{I< J} K_{IJ}}$ in Eq.~\eqref{1SPTinv} to obtain
\begin{equation}\label{1SPTinv0}
\begin{aligned}
\hspace{-5pt}Z^\text{top}(\cM^4, B^\QZ_I\hspace{-1pt}) &\hspace{-2pt}=\hspace{-2pt} 
\ee^{\ii \pi \hspace{-2pt}\sum\limits_{I, J} \hspace{-2pt}K_{IJ} \hspace{-4pt}\int\limits_{\cM^4} \hspace{-2pt} B_I^\QZ  B_J^\QZ + B^{\QZ}_I \hcup{1} \dd  B^{\QZ}_J}  \\
&\ \ \ \ \times\ee^{\ii \pi  
\hspace{-2pt}\sum\limits_{I<J} \hspace{-2pt}K_{IJ} \hspace{-4pt}\int\limits_{\cM^4} \hspace{-2pt} \dd B_I^\QZ \hcup{2} \dd B_J^\QZ
} .
\end{aligned}
\end{equation}
Eq.~\eqref{1SPTinv0} thus provides a form of the 1-SPT invariant in terms of only the $K$-matrix. However, due to the sum on the second term only being over ${I<J}$, it is not covariant.

We see that, from Eq.~\eqref{1SPTinv}, the 1-SPT invariant is characterized by
a pair of integer matrices $(K,k)$. At first glance, due to the $k$ dependence,
or equivalently Eq.~\eqref{1SPTinv0} not being covariant, it appears that the
SPT invariant is changed by the transformation ${B^{\Q/\Z}_I \to \t B^{\Q/\Z}_I
}$ and ${K\to \t K }$ where
\begin{align}
\label{BKtK}
\t B^{\Q/\Z}_I = (U^{-1})_{IJ} B^{\Q/\Z}_J,\ \ \ 
\t K =  U^\top K U ,
\end{align}
and ${U_{IJ} \in GL(\ka,\Z)}$. Therefore, it would appear that $K$
and $\t K$ do not describe the same 1-SPT invariant.

However, it turns out that $K$ and $\t K$ actually do describe the same 1-SPT invariant. To show this, we first show that the 1-SPT invariant is left unchanged when
$k$ is replaced by another integer matrix $k'$ (not necessarily upper triangular) such that
${K = k'+k'^\top}$.  The difference ${A=k-k'}$ is an antisymmetric integer matrix.
The respective lattice Lagrangian densities of the 1-SPT invariant Eq.~\eqref{1SPTinv} for $k$ and $k'$
(after dividing by $2\pi$) differ by
\begin{equation*}
\sum_{I, J} \frac{A_{IJ}}{2}  \dd B_I^\QZ \hcup{2} \dd B_J^\QZ.
\end{equation*}
Using that $A$ is antisymmetric, that integer multiples of $2\pi$ can be added to the Lagrangian density without changing the path integral, and Eq.~\eq{cupkrel}, this can be rewritten as
\begin{equation*}
\begin{aligned}
&\sum_{I, J} \frac{A_{IJ}}{2}  \dd B_I^\QZ \hcup{2} \dd B_J^\QZ\\
&\hspace{20pt}\se{1}\sum_{I< J} \frac{A_{IJ}}{2} \dd \big(-\dd B_I^\QZ \hcup{3} \dd B_J^\QZ\big),
\end{aligned}
\end{equation*}
which vanishes on a closed manifold. Therefore, the two Lagrangian densities differ
by only a coboundary term and give the same topological invariants for a closed
$\cM^4$.
 
The above result allows us to show that the
SPT invariant is  unchanged under the transformation Eq.~\eqref{BKtK}.  Indeed, we now only need to check the $k_{IJ}$ term in
Eq.~\eqref{1SPTinv}.  Let $\bar k$ be an integer matrix defined by
\begin{equation}
 \bar k_{IJ} = \sum_{I'\leq J'} (U^\top)_{II'} k_{I'J'} U_{J'J},
\end{equation}
such that ${\t K = \bar k + \bar k^\top}$. Using that, from Eq.~\eqref{BKtK}, ${B^{\Q/\Z}_I = U_{IJ} \t B^{\Q/\Z}_J}$ and plugging it into the second line of Eq.~\eqref{1SPTinv}, it becomes
\begin{equation}
\ee^{\ii \pi  \hspace{-2pt}\sum\limits_{I\leq J} \hspace{-2pt}k_{IJ} \hspace{-4pt}\int\limits_{\cM^4} \hspace{-2pt} \dd B_I^\QZ \hcup{2} \dd B_J^\QZ } \hspace{-8pt}= \ee^{\ii \pi  \hspace{-2pt}\sum\limits_{I, J} \hspace{-1 pt} \bar k_{IJ} \hspace{-4pt}\int\limits_{\cM^4} \hspace{-2pt}\dd \t B_I^\QZ \hcup{2} \dd \t B_J^\QZ } .
\end{equation}
Let's now introduce the upper triangular integer matrix $\t k$ such that ${\t K = \t k + \t k^\top}$. Using the above result for a closed $\cM^4$, the SPT invariant is unchanged by replacing $\bar k$ with $\t k$. Therefore, the SPT invariant Eq.~\eqref{1SPTinv} is unchanged under the transformation Eq.~\eqref{BKtK}.

The fact that $K$ and $\t K$ describes the same SPT invariant also allows us to find a convenient expression for the SPT invariant. 
Indeed, recall that the integer matrix $K$ has the Smith normal form given by Eq.~\eqref{smith}. From the above discussion, we see that,
without loosing generality, we may transform ${K\to U^\top K U}$ without changing the SPT invariant and thus may assume $K$ to have the following form
\begin{align}\label{kDecomp}
K= 
V^\top
 \begin{pmatrix}
 k_1 & & & \\
  &  k_2 & & \\
  &   &   k_3 & \\
  &   &    & \ddots \\
\end{pmatrix} 
=
 \begin{pmatrix}
 k_1 & & & \\
  &  k_2 & & \\
  &   &   k_3 & \\
  &   &    & \ddots \\
\end{pmatrix} V .
\end{align}
The invertible integer matrix $V$ satisfies
\begin{align}
 (V^\top)_{IJ} k_J = k_J V_{JI} =  k_I V_{IJ} 
\ \ \text{ or } \ \
\frac{V_{IJ}}{V_{JI}} = \frac{k_J}{k_I}.
\end{align}
Using this expression for $K$, the 1-SPT invariant in its original form given by Eq.~\eqref{SPTinvkIJ} can be rewritten as
\begin{equation*}
\begin{aligned}
&
Z^\text{top}(\cM^4, B^\QZ_I) = \ee^{\ii \pi \sum\limits_{I} k_{I}V_{II} 
\hspace{-4pt}\int\limits_{\cM^4} \hspace{-2pt}  B_I^\QZ  B_I^\QZ
+  B^{\QZ}_I \hcup{1} \dd  B^{\QZ}_I } \times \\
&
\hspace{80pt} \ee^{\ii 2\pi \sum\limits_{I< J} \hspace{-1pt}k_{I}V_{IJ} 
\hspace{-4pt}\int\limits_{\cM^4} \hspace{-2pt}  B_J^\QZ  B_I^\QZ
+  B^{\QZ}_I \hcup{1} \dd  B^{\QZ}_J } .
\end{aligned}
\end{equation*}
Furthermore, the quantization condition on the background cochain field then becomes ${\sum_I B^\QZ_{I} k_I V_{IJ} \se{1}0}$. This can be automatically satisfied if we let $B^\QZ_{I}$ take the form
\begin{align}
 B^\QZ_{I} = k_I^{-1} B^{\Z_{k_I}}_I,
\end{align}
where $B^{\Z_{k_I}}_I$ is a $\Z_{k_I}$-valued 2-cocycle and thus satisfies ${\dd  B^{\Z_{k_I}}_I \se{k_I} 0 }$. Using this, the 1-SPT invariant for the ${\Z_{k_1}^{(1)}\times \Z_{k_2}^{(1)}\times \cdots}$ 1-symmetry becomes
\begin{align}
\label{1SPTinvk}
&
Z^\text{top}(\cM^4, B^{\Z_{k_I}}_I) = \ee^{\ii \pi \sum\limits_{I} V_{II}k_I^{-1} 
\hspace{-4pt}\int\limits_{\cM^4} \hspace{-2pt}   B_I^{\Z_{k_I}}  B_I^{\Z_{k_I}}
+  B^{\Z_{k_I}}_I \hcup{1} \dd  B^{\Z_{k_I}}_I } 
\nonumber \\
&\hspace{30pt}\times
 \ee^{\ii 2\pi \sum\limits_{I< J}\hspace{-1pt} V_{IJ}k_J^{-1} 
\hspace{-4pt}\int\limits_{\cM^4} \hspace{-2pt}    B_J^{\Z_{k_J}}  B_I^{\Z_{k_I}}
+  B^{\Z_{k_I}}_I \hcup{1} \dd  B^{\Z_{k_J}}_J }
.
\end{align}

\subsection{Some Examples of SPT Invariants}\label{examplesSec}

In the previous section, we found that $U^\kappa(1)$ gauge theory with a $2\pi$
topological term in the confined phase has a non-trivial 1-SPT invariant,
Eq.~\eqref{SPTinvkIJ}. We then massaged the SPT invariant into other forms,
such as Eq.~\eqref{1SPTinv0} and Eq.~\eqref{1SPTinvk}. This suggests that
generically there is a phase in the confined phase with non-trivial 1-SPT order
which is protected by the ${\Z_{k_1}^{(1)}\times \Z_{k_2}^{(1)}\times \cdots}$
symmetry discussed in section~\ref{sec:syms}. Now we will consider same simple
examples of different $K$-matrices and the corresponding 1-SPT order. The first
example will have ${\kappa = 1}$ while the second and third will be ${\kappa =
2}$.

\vspace{10pt}
\begin{center}
\textit{Example 1}
\end{center}
\vspace{10pt}

Let's first consider the case where there is only one type of cochain field $a^\RZ$ so ${\ka=1}$ and the $K$-matrix would become 
\begin{equation}
K=
\begin{pmatrix} 
2n 
\end{pmatrix},
\end{equation}
with ${n\in\Z}$. In this case, the 3+1D bosonic model on spacetime lattice Eq.~\eqref{K4Ddada} becomes
\begin{align}
Z = \int D[a^\RZ] & \
\ee^{-  \int_{\cM^4} \frac{|\dd a^\RZ-\toZ{\dd a^\RZ}|^2}{g} }
 \\
&
 \ee^{\ii 2\pi n \int_{\cM^4} (\dd a^\RZ-\toZ{\dd a^\RZ})(\dd a^\RZ-\toZ{\dd a^\RZ})}
\nonumber \\
&
\ee^{\ii 4\pi n \int_{\cM^4}   a^{\RZ} \dd \toZ{\dd a^{\RZ}} 
-  \dd a^{\RZ} \hcup{1}\dd \toZ{\dd a^{\RZ}} }.
\nonumber 
\end{align}
From our previous discussion, this theory has a $\Z_{2n}^{(1)}$ symmetry. Let's see this explicitly. The path integral is invariant under the transformation ${a^\RZ \to a^\RZ + \frac1{2n}\beta^\Z}$ where $\beta^\Z$ is an arbitrary $\Z$-valued 1-cochain satisfying ${\dd\beta^\Z \se{2n} 0}$. The physical part of $\beta^\Z$ is defined modulo $2n$ because shifting $\beta^\Z$ by $2n$-valued 1-cochain corresponds to shifting $a^\RZ$ by an integer-valued 1-cochain, which is a gauge transformation. Therefore, this theory indeed has a $\Z_{2n}^{(1)}$ symmetry. When $g\ll 1$, the above bosonic model at low-energies describes the deconfined phase of $U(1)$ gauge field theory. At energies much smaller than the energy gap of the $U(1)$ monopole, $\dd{\toZ{\dd a^\RZ} = 0}$ and the $\Z_{2n}^{(1)}$
symmetry is promoted to an emergent $U(1)^{(1)}$ symmetry.

When $g\gg 1$, the above bosonic model is in a gapped phase with
$\Z_{2n}^{(1)}$ 1-symmetry, which corresponds to the confined phase of the
$U(1)$ gauge theory. From our general discussion, this gapped phase is an SPT phase protected by the $\Z_{2n}^{(1)}$ 1-symmetry.  Indeed, using Eq.~\eq{1SPTinvk}, this SPT phase is characterized by the 1-SPT invariant
\begin{equation}
\begin{aligned}
Z^\text{top}(\cM^4, B^{\Z_{2n}}) 
&= \ee^{\ii \frac{2\pi}{4n} \hspace{-4pt}\int\limits_{\cM^4} \hspace{-2pt}   B^{\Z_{2n}}  
B^{\Z_{2n}} + B^{\Z_{2n}} \hcup{1} \dd  B^{\Z_{2n}}},
\\
&= \ee^{\ii \frac{2\pi}{4n} \hspace{-4pt}\int\limits_{\cM^4} \hspace{-2pt} \gSq^2(B^{\Z_{2n}}) }  
.
\end{aligned}
\end{equation}
The 3+1D 1-SPT order for the $\Z_{2n}^{(1)}$ 1-symmetry is classified by
${H^4(B(\Z_{2n},2);\RZ)=\Z_{4n}}$.\cite{TW190802613}  From the SPT invariant,
we find that the 1-SPT order realized by the confined phase is given by $1\in
\Z_{4n}$, and thus is the generator of the SPT orders classified by
$H^4(B(\Z_{2n},2);\RZ)$.

\vspace{10pt}
\begin{center}
\textit{Example 2}
\end{center}
\vspace{10pt}

Let's now consider an example where there are two types of 1-cochain fields $a^\RZ_1$ and $a^\RZ_2$, so ${\kappa = 2}$, and the $K$-matrix is given by
\begin{equation}\label{ex2K1}
K = 
\begin{pmatrix} 
2 & 1\\ 
1& 2\\ 
\end{pmatrix}.
\end{equation}
We'd like to find the SPT invariant for this $K$ matrix using Eq.~\eq{1SPTinvk}. This requires us to first find the integers ${k_1,k_2}$ and the integer matrix $V$ from Eq.~\eqref{kDecomp}. The diagonal elements of the Smith normal form of $K$ are ${(k_1,k_2) = (3,1)}$. Thus, by finding $k_I$ we can immediately conclude that the 1-symmetry is ${\Z_3^{(1)}\times \Z_1^{(1)}\equiv\Z_3^{(1)}}$. However, there does not exist an integer matrix $V$ which will work for this $K$. 

To find the SPT invariant, we can instead consider the matrix
\begin{equation}
\t K = 
\begin{pmatrix} 
6 & 3\\ 
3 & 2\\ 
\end{pmatrix}.
\end{equation}
Since ${\t K = U K U^\top}$, where
\begin{equation}
U =  
\begin{pmatrix} 
1 & 1\\ 
1& 0\\ 
\end{pmatrix}
\in GL(2,\Z),
\end{equation}
our results from section~\ref{section:SPTinv} show that the SPT order of $\t K$ is equivalent to that of the $K$ matrix Eq.~\eqref{ex2K1}. Therefore, we now attempt to find the SPT invariant using the same approach but now with $\t K$. First note that the diagonal elements of the Smith normal form of $\t K$ are still ${(k_1,k_2) = (3,1)}$. The $\t K$ matrix can be written as
\begin{align}
\t K = 
\begin{pmatrix} 
6 & 3\\ 
3 & 2\\ 
\end{pmatrix}
= 
\begin{pmatrix} 
3 & 0\\ 
0 & 1\\ 
\end{pmatrix}
\begin{pmatrix} 
2 & 1\\ 
3 & 2\\ 
\end{pmatrix},
\end{align}
and we find the integer matrix $V$ to be
\begin{equation*}
V=\begin{pmatrix} 
2 & 1\\ 
3 & 2\\ 
\end{pmatrix}.
\end{equation*}

From the $k_I$ and $V$ found for $\t K$, we have that
\begin{align}
(V_{IJ}k_J^{-1})  =
\begin{pmatrix} 
\frac23 & 1\\ 
1 & 2\\ 
\end{pmatrix} .
\end{align}
Using Eq.~\eqref{1SPTinvk}, the corresponding 1-SPT invariant for the ${\Z_3^{(1)}\times \Z_1^{(1)}\equiv\Z_3^{(1)}}$ 1-symmetry is given by 
\begin{equation}
\begin{aligned}
&
Z^\text{top}(\cM^4, B^{\Z_{k_I}}_I) = 
\ee^{\ii \frac{2\pi}{3} \hspace{-4pt}\int\limits_{\cM^4} \hspace{-2pt}
  B_1^{\Z_3} B_1^{\Z_3} + B^{\Z_3}_1 \hcup{1} \dd  B^{\Z_3}_1} \times\\
&\hspace{90pt}
\ee^{\ii 2\pi\hspace{-4pt}\int\limits_{\cM^4} \hspace{-2pt} B_2^{\Z_1} B_1^{\Z_3} + B^{\Z_3}_1 \hcup{1} \dd  B^{\Z_1}_2}\times\\
&\hspace{90pt}
\ee^{\ii 2\pi \hspace{-4pt}\int\limits_{\cM^4} \hspace{-2pt}
  B_2^{\Z_1} B_2^{\Z_1} + B^{\Z_1}_2 \hcup{1} \dd  B^{\Z_1}_2}.
\end{aligned}
\end{equation}
We can now use the fact that the SPT invariant is invariant under the gauge transformation ${B_2^{\Z_1} \to B_2^{\Z_1} + m^\Z}$, where $m^\Z$ is a $\Z$-valued 2-cochain, to set ${B_2^{\Z_1} = 0}$. Doing so, the SPT invariant simplifies to
\begin{equation}\label{example2SPTinv}
Z^\text{top}(\cM^4, B^{\Z_{k_I}}_I) = 
\ee^{\ii \frac{2\pi}{3} \hspace{-4pt}\int\limits_{\cM^4} \hspace{-2pt}
  B_1^{\Z_3} B_1^{\Z_3} + B^{\Z_3}_1 \hcup{1} \dd  B^{\Z_3}_1}.
\end{equation}
Therefore, the 1-SPT invariant of the K matrix Eq.~\eqref{ex2K1} is given by
Eq.~\eqref{example2SPTinv}. 1-SPT order protected by the 1-symmetry
$\Z_3^{(1)}$ is classified by ${H^4(B(\Z_3,2);\RZ)=\Z_3}$.\cite{TW190802613}
Therefore, from Eq.~\eqref{example2SPTinv} the SPT order realized in the
confined phase for the $K$-matrix Eq.~\eqref{ex2K1} is given by $1 \in \Z_3$
and is thus the generator for SPT orders classified by  ${H^4(B(\Z_3,2);\RZ)}$.

\vspace{10pt}
\begin{center}
\textit{Example 3}
\end{center}
\vspace{10pt}

For our final example, let's again consider the scenario where there are ${\kappa = 2}$ cochain fields $a^\QZ_I$, but now where the $K$-matrix is
\begin{equation}\label{Kmatrxex3}
K=
\begin{pmatrix} 
0 & n\\ 
n& 0\\ 
\end{pmatrix}, \ \ \ \ n\in\Z.
\end{equation}
The diagonal elements of the Smith normal form of this $K$-matrix are
${(k_1,k_2) = (n,n)}$. Therefore, the model with this $K$ matrix has a
${\Z_n^{(1)}\times \Z_n^{(1)}}$ symmetry. Furthermore, this $K$ matrix can be
written as
\begin{align}
K = 
\begin{pmatrix} 
0 & n\\ 
n & 0\\ 
\end{pmatrix}
= 
\begin{pmatrix} 
n & 0\\ 
0 & n\\ 
\end{pmatrix}
\begin{pmatrix} 
0 & 1\\ 
1 & 0\\ 
\end{pmatrix} .
\end{align}
Thus, unlike example 2, using the $K$ matrix we start with, there exists the
integer matrix 
\begin{equation}
V=\begin{pmatrix} 
0 & 1\\ 
1 & 0\\ 
\end{pmatrix}.
\end{equation}
From this matrix $V$ and from $k_I$, we find that
\begin{equation}
(V_{IJ}k_J^{-1})  =
\begin{pmatrix} 
0 & \frac1n\\ 
\frac1n & 0\\ 
\end{pmatrix} .
\end{equation}

Using Eq.~\eqref{1SPTinvk}, the corresponding 1-SPT invariant for the $\Z_n^{(1)}\times \Z_n^{(1)}$ 1-symmetry is given by 
\begin{align}
Z^\text{top}(\cM^4, B^{\Z_{n}}_I) 
= 
\ee^{\ii \frac{2\pi}n \int_{\cM^4}  
  B_2^{\Z_n} B_1^{\Z_n} + B^{\Z_n}_1 \hcup{1} \dd  B^{\Z_n}_2}
.
\end{align}
Thus, we find that the 1-SPT order in the confined phase of ${U(1)\times U(1)}$
3+1D gauge theory with $K$ matrix Eq.~\eqref{Kmatrxex3} is a mixed SPT order
between the two $\Z_n^{(1)}$ 1-symmetries. In other words, the boundary
Chern-Simons theory has a mixed anomaly between two $\Z_n^{(1)}$ 1-symmetries.
This Chern-Simons theory describes 2+1D $\Z_n$ topological order. Indeed, the
loop operators charged under the two $\Z_n^{(1)}$ 1-symmetries are the loop
objects whose open ends correspond to $e$ and $m$ type anyons, respectively.
Furthermore, the fact that the $e$ and $m$ anyons have nontrivial mutual
statistics is a manifestation of the mixed anomaly between the two $\Z_n^{(1)}$
1-symmetries.

\section{Conclusion}\label{section:conclusion}

In this paper, we have considered 3+1D compact $U^\ka(1)$ gauge theory with
$2\pi$-quantized topological terms. In section~\ref{latticemodel}, we developed
a bosonic lattice model acting as the UV regularization for the continuum
theory.  Working with this lattice model, we found that at energies much
smaller than the gauge charges' gaps but larger than the monopoles' gaps, there
is an exact emergent ${\Z_{k_1}^{(1)} \times \Z_{k_2}^{(1)} \times \cdots}$
1-symmetry. We found that the confined phase of the $U^\ka(1)$ gauge theory
(\ie the symmetric gapped phase of the bosonic model) has non trivial symmetry
protected topological (SPT) order which is protected by the exact emergent
${\Z_{k_1}^{(1)} \times \Z_{k_2}^{(1)} \times \cdots}$ 1-symmetry. We then went
on to gauge this symmetry in section~\ref{gaugeSym} and found the
corresponding SPT invariant in section~\ref{section:SPTinv}.  We gave some
examples of different $K$ matrices where the confined phases of the $U^\ka(1)$ gauge theories realizes a
$\Z_{2n}^{(1)}$ 1-SPT phase, a $\Z_{3}^{(1)}$ 1-SPT phase, and a
$\Z_{n}^{(1)}\times \Z_{n}^{(1)}$ mixed 1-SPT phase.

\textit{Note}: after the completion of this paper, we noticed the independent work \Rf{MF220607725} which studied the emergent 1-symmetry for the ${\ka=1}$ case in a phase where monopoles were only partially condensed.

\section{Acknowledgements}

SDP is thankful for helpful and fun discussions with Hart Goldman, Ethan Lake and Ho Tat Lam on higher-form symmetries and with Michael DeMarco on lattice Chern-Simons theory. SDP, additionally, acknowledges support from the Henry W. Kendall
Fellowship. This work is partially supported by NSF DMR-2022428 and by the Simons Collaboration on Ultra-Quantum Matter, which is a grant from the Simons Foundation (651446, XGW). 

\appendix
\allowdisplaybreaks

\section{The Gauge Invariance of the 1-SPT Invariant}
\label{BgaugeInv}

In section~\ref{section:SPTinv} of the main text, we found that the 1-SPT invariant for the $\Z_{k_1}^{(1)} \times \Z_{k_2}^{(1)} \times \cdots
$ 1-symmetry given by Eq.~\eqref{SPTinvkIJ}:
\begin{equation*}
Z^\text{top}(\cM^4, B^\QZ_I) = 
 \ee^{\ii 2\pi \hspace{-2pt}\sum\limits_{I\leq J} \hspace{-2pt}k_{IJ} \hspace{-4pt}\int\limits_{\cM^4} \hspace{-2pt}  B_J^\QZ  B_I^\QZ
+  B^{\QZ}_I \hcup{1} \dd  B^{\QZ}_J }.
\end{equation*}
Here, $B^{\QZ}_I$, with ${I=1,\ldots\kappa}$, are background symmetry twist 2-cochain fields satisfying
\begin{equation*}
\dd  B^{\QZ}_I \se{1} 0, \ \ \ \ \
\sum_I B^\QZ_{I} K_{IJ} \se{1}0.
\end{equation*}
In this appendix section, we confirm a claim made in the main text that the above 1-SPT invariant for closed $\cM^4$ is invariant under
the gauge transformations
\begin{align}
\label{Bgauge}
 B^{\QZ}_I \to B^{\QZ}_I + n_I,\ \ \ \ \
 B^{\QZ}_I \to B^{\QZ}_I + \dd a_I^\QZ,
\end{align}
where $n_I$ are $\Z$-valued 2-cochains and $a_I^\QZ$ are $\QZ$-valued
1-cochains satisfying the quantization conditions ${\sum_I a^\QZ_{I} K_{IJ}
\se{1}0}$.  

First, we'll check the $\Z$-gauge transformation ${B^{\QZ}_I \to B^{\QZ}_I + n_I}$, which causes the 1-SPT invariant to change by a factor 
\begin{equation*}
\ee^{\ii 2\pi \hspace{-2pt}\sum\limits_{I\leq J} \hspace{-2pt}k_{IJ} \hspace{-4pt}\int\limits_{\cM^4} \hspace{-2pt} n_J  B_I^\QZ +B_J^\QZ  n_I }\hspace{-2pt}\ee^{\ii 2\pi \hspace{-2pt}\sum\limits_{I\leq J} \hspace{-2pt}k_{IJ} \hspace{-4pt}\int\limits_{\cM^4} \hspace{-2pt} n_I \hcup{1} \dd  B^{\QZ}_J +B^{\QZ}_I \hcup{1} \dd  n_J }\hspace{-2pt}.
\end{equation*}
Assuming that $\prt \cM^4 =\emptyset$ and using
\eqref{cupkrel}, this can be rewritten as unity:
\begin{equation*}
\begin{aligned}
&\ee^{\ii 2\pi \hspace{-2pt}\sum\limits_{I\leq J} \hspace{-2pt}k_{IJ} \hspace{-4pt}\int\limits_{\cM^4} \hspace{-2pt} n_J  B_I^\QZ +B_J^\QZ  n_I }\hspace{-2pt}\ee^{\ii 2\pi \hspace{-2pt}\sum\limits_{I\leq J} \hspace{-2pt}k_{IJ} \hspace{-4pt}\int\limits_{\cM^4} \hspace{-2pt} n_I \hcup{1} \dd  B^{\QZ}_J +B^{\QZ}_I \hcup{1} \dd  n_J }\hspace{-2pt}\\
&=
 \ee^{\ii 2\pi \hspace{-2pt}\sum\limits_{I, J} \hspace{-2pt}K_{IJ} \hspace{-4pt}\int\limits_{\cM^4} \hspace{-2pt}B_J^\QZ  n_I }
 \ee^{\ii 2\pi \hspace{-2pt}\sum\limits_{I\leq J} \hspace{-2pt}k_{IJ} \hspace{-4pt}\int\limits_{\cM^4} \hspace{-2pt}   n_J B_I^\QZ - B_I^\QZ  n_J }
\nonumber \\
&\ \ \ \ \ \quad \times\ee^{\ii 2\pi \hspace{-2pt}\sum\limits_{I\leq J} \hspace{-2pt}k_{IJ} \hspace{-4pt}\int\limits_{\cM^4} \hspace{-2pt} n_I \hcup{1} \dd  B^{\QZ}_J +B^{\QZ}_I \hcup{1} \dd  n_J }
\nonumber\\
&=
 \ee^{\ii 2\pi \hspace{-2pt}\sum\limits_{I\leq J} \hspace{-2pt}k_{IJ} \hspace{-4pt}\int\limits_{\cM^4} \hspace{-2pt}   - \dd B_I^\QZ \hcup{1} n_J - B_I^\QZ \hcup{1} \dd n_J }
\nonumber \\
&\ \ \ \ \quad \times\ee^{\ii 2\pi \hspace{-2pt}\sum\limits_{I\leq J} \hspace{-2pt}k_{IJ} \hspace{-4pt}\int\limits_{\cM^4} \hspace{-2pt} n_I \hcup{1} \dd  B^{\QZ}_J +B^{\QZ}_I \hcup{1} \dd  n_J }
\nonumber\\
&= \ee^{\ii 2\pi \hspace{-2pt}\sum\limits_{I\leq J} \hspace{-2pt}k_{IJ} \hspace{-4pt}\int\limits_{\cM^4} \hspace{-2pt}   n_I \hcup{1} \dd  B^{\QZ}_J  - \dd B_I^\QZ \hcup{1} n_J  }
\nonumber\\
&= \ee^{\ii 2\pi \hspace{-2pt}\sum\limits_{I\leq J} \hspace{-2pt}k_{IJ} \hspace{-4pt}\int\limits_{\cM^4} \hspace{-2pt}   n_I \hcup{1} \dd  B^{\QZ}_J  + n_J \hcup{1} \dd B_I^\QZ  + \dd n_J \hcup{2} \dd B_I^\QZ  }
\nonumber\\
&= \ee^{\ii 2\pi \hspace{-2pt}\sum\limits_{I, J} \hspace{-2pt}K_{IJ} \hspace{-4pt}\int\limits_{\cM^4} \hspace{-2pt}  n_I \hcup{1} \dd  B^{\QZ}_J      } = 1.
\end{aligned}
\end{equation*}
Therefore, the SPT invariant is unchanged by the gauge transformation ${B^{\QZ}_I \to B^{\QZ}_I + n_I}$.

Lastly, let's check the gauge transformation ${B^{\QZ}_I \to B^{\QZ}_I + \dd a_I^\QZ}$, which causes the 1-SPT invariant to change by a factor
\begin{equation*}
\ee^{\ii 2\pi \hspace{-2pt}\sum\limits_{I\leq J} \hspace{-2pt}k_{IJ} \hspace{-4pt}\int\limits_{\cM^4} \hspace{-2pt} \dd a_J^\QZ  B_I^\QZ 
+B_J^\QZ  \dd a_I^\QZ 
+ \dd a_I^\QZ \hcup{1} \dd  B^{\QZ}_J  }.
\end{equation*}
Once again, assuming $\prt \cM^4 =\emptyset$ and using
\eqref{cupkrel}, we can show that this change is equal to unity:
\begin{align}
&\ee^{\ii 2\pi \hspace{-2pt}\sum\limits_{I\leq J} \hspace{-2pt}k_{IJ} \hspace{-4pt}\int\limits_{\cM^4} \hspace{-2pt} \dd a_J^\QZ  B_I^\QZ 
+B_J^\QZ  \dd a_I^\QZ 
+ \dd a_I^\QZ \hcup{1} \dd  B^{\QZ}_J  }
\nonumber\\
&=
\ee^{\ii 2\pi \hspace{-2pt}\sum\limits_{I\leq J} \hspace{-2pt}k_{IJ} \hspace{-4pt}\int\limits_{\cM^4} \hspace{-2pt} \dd a_J^\QZ  B_I^\QZ 
+   \dd a_I^\QZ B_J^\QZ 
 - \dd a_I^\QZ  \hcup{1} \dd B_J^\QZ 
}
\nonumber\\
&\ \ \ \ \quad\times
\ee^{\ii 2\pi \hspace{-2pt}\sum\limits_{I\leq J} \hspace{-2pt}k_{IJ} \hspace{-4pt}\int\limits_{\cM^4} \hspace{-2pt}
 \dd a_I^\QZ \hcup{1} \dd  B^{\QZ}_J  }
\nonumber\\
&=\ee^{\ii 2\pi \hspace{-2pt}\sum\limits_{I, J} \hspace{-2pt}K_{IJ} \hspace{-4pt}\int\limits_{\cM^4} \hspace{-2pt}\dd a_J^\QZ  B_I^\QZ} =1
\end{align}
Therefore, the SPT invariant is also unchanged by the gauge transformation ${B^{\QZ}_I \to B^{\QZ}_I + \dd a_I^\QZ}$.

\bibliography{
../../bib/all,
../../bib/allnew,
../../bib/publst,
../../bib/publstnew,
references,
../../bib/salRefs
}
\end{document}